\documentclass[a4paper,useAMS,usenatbib]{mnras}
\pdfoutput=1 

\usepackage{graphicx}
\usepackage{setspace}
\usepackage{natbib}
\usepackage{color}
\usepackage{amsmath,amssymb}
\usepackage{times}
\usepackage{float}

\usepackage{hyperref}[]

\title[In-situ vs accreted MW globular clusters]{In-situ vs accreted Milky Way globular clusters: a new classification method and implications for cluster formation}

\author[Belokurov \& Kravtsov]{Vasily  Belokurov$^{1}$\thanks{E-mail:vasily@ast.cam.ac.uk} and Andrey Kravtsov$^{2,3,4}$\thanks{E-mail:kravtsov@uchicago.edu}\\
  $^1$Institute of Astronomy, Madingley Rd, Cambridge, CB3 0HA, UK\\ 
  $^2$Department of Astronomy and Astrophysics, The University of Chicago, Chicago, IL 60637 USA\\
  $^3$Kavli Institute for Cosmological Physics, The University of Chicago, Chicago, IL 60637 USA\\
  $^4$Enrico Fermi Institute, The University of Chicago, Chicago, IL 60637}

\begin{document}
\defcitealias{Aurora}{BK22}

\maketitle

\label{firstpage}

\begin{abstract}
We present a new scheme for the classification of the in-situ and accreted globular clusters (GCs). The scheme uses total energy $E$ and $z$-component of the orbital angular momentum and is calibrated using [Al/Fe] abundance ratio. We demonstrate that such classification results in the GC populations with distinct spatial, kinematic, and chemical abundance distributions.
The in-situ GCs are distributed within the central 10 kpc of the Galaxy in a flattened configuration aligned with the MW disc, while the accreted GCs have a wide distribution of distances and a spatial distribution close to spherical. In-situ and accreted GCs have different $\rm [Fe/H]$ distributions with the well-known bimodality present only in the metallicity distribution of the in-situ GCs. Furthermore, the accreted and in-situ GCs are well separated in the plane of $\rm [Al/Fe]-[Mg/Fe]$ abundance ratios and follow distinct sequences in the age--$\rm [Fe/H]$ plane.  The in-situ GCs in our classification show a clear disc spin-up signature -- the increase of median $V_\phi$ at metallicities $\rm [Fe/H]\approx -1.3\div -1$ similar to the spin-up in the in-situ field stars. This signature signals the MW's disc formation, which occurred $\approx 11.7-12.7$ Gyrs ago (or at $z\approx 3.1-5.3$) according to GC ages. In-situ GCs with metallicities of $\rm [Fe/H]\gtrsim -1.3$ were thus born in the Milky Way disc, while lower metallicity in-situ GCs were born during early, turbulent, pre-disc stages of the evolution of the Galaxy and are part of its Aurora stellar component.
\end{abstract}


\section{Introduction}
\label{sec:intro}

The Milky Way offers an uninterrupted view of the time evolution of a single galaxy, thus providing us with a useful benchmark for the theory of galaxy formation \citep[e.g., see][for a review]{Bland_Hawthorn.Gerhard.2016}. In the hierarchical structure formation scenario \citep{Peebles.1965,Peebles.Yu.1970} galaxy evolution is driven by the formation of in-situ stars in the main progenitor \citep{Eggen.etal.1962} and accretion of stars from the smaller galaxies that merge with it \citep{Searle.1977}. Globular clusters (GCs) have long been used to elucidate the early phases of the Milky Way's formation, in particular the relative importance of the in-situ formation and the accretion of sub-galactic fragments \citep[e.g.,][]{Searle.Zinn.1978,Cote.etal.2000,Forbes2010,Leaman2013,Massari2019}. However, the origin of clusters themselves was until recently rather uncertain, with ideas of their formation spanning from the Jeans fragmentation in the early intergalactic medium \citep{Peebles.Dicke.1968}, formation predominantly in the cores of dwarf galaxies \citep{Peebles.1984,Searle.Zinn.1978}, thermal instability in the halo gas of the MW progenitor \citep{Fall.Rees.1985}, gas compressions due to shocks in the primordial molecular clouds \citep{Gunn.1980,Murray.Lin.1992,Harris.Pudritz.1994,Burkert.etal.1996}, gas compression produced during major mergers \citep{Schweizer.1987,Ashman.Zepf.1992,Ashman.Zepf.2001}.

Our understanding of globular cluster formation was revolutionized by the high-resolution observations with the Hubble Space Telescope (HST). Early HST observations confirmed efficient formation of compact GC-like objects in merging galaxies during their final galaxy collision stage \citep[e.g.,][]{Whitmore.etal.1993,Whitmore.etal.1999,Whitmore.Schweizer.1995,Holzman.etal.1996,Zepf.etal.1999}. However, subsequent observations of a wider range of galaxies showed that globular clusters form as part of regular star formation in galaxies where gas and star formation surface densities are sufficiently large \citep[e.g., see][for reviews]{Krumholz2019,Adamo.etal.2020rev}.

Indeed, models in which GC formation was implemented as part of a regular star formation during gas-rich phases of galaxy evolution in the hierarchical cosmological framework proved quite successful in matching observed properties of GC populations \citep[e.g.,][]{Cote.etal.2000,Cote.etal.2002,Beasley.etal.2002,Kravtsov2005,Muratov.Gnedin.2010,Kruijssen2015,Choksi2018,Choksi.Gnedin.2019,Kruijssen.etal.2019,Chen_Gnedin2022,Chen_Gnedin2023,Reina.Campos.etal.2022}. It is thus now generally acknowledged that GCs are tracing star formation in galaxies, albeit at specific epochs when conditions conducive for their formation exist \citep[e.g.,][]{Kruijssen2015,Choksi2018,Reina.Campos.etal.2022}. Thus, for example, the Milky Way has not been forming globular clusters for the past $\sim 8-10$ Gyrs, while it formed most of its in-situ stellar population during these epochs. 

Furthermore, the number of GCs scales approximately linearly with halo mass \citep[see][]{Spitler.Forbes.2009,Hudson.etal.2014,Harris.etal.2017,Forbes.etal.2018,Dornan.Harris.2023}, while stellar mass of galaxies with luminosities $L\lesssim L_\star$ scales much faster, $M_\star\propto M_{\rm h}^\alpha$ where $\alpha\approx 1.5-2.5$ depending on $M_\star$ \citep[e.g.,][]{Kravtsov.etal.2018,Nadler.etal.2020}. This means that the number of GCs per stellar mass increases with decreasing $M_\star$ and accreted dwarf galaxies contribute proportionally more GCs to the host galaxy than stars. 

In the Milky Way
globular cluster formation is biased towards earlier epochs when galaxies experienced larger accretion and merger rates and were generally considerably more gas-rich. GCs can thus be a useful probe of the Galaxy evolution and merger history during these early epochs.
However, this requires a way to differentiate GCs that were born in-situ in the main MW progenitor and GCs that were accreted as part of other galaxies. The earliest efforts to identify accreted and in-situ clusters were based on the metallicity and spatial distribution of GCs. \citet[][see also \citealt{Zinn.1996} for review]{Zinn.1985} divided clusters by metallicity at $\rm [Fe/H]=-0.8$ and argued that such division resulted in GC populations with distinct spatial and kinematic properties. For example, it was claimed that the metal-richer Galactic GCs likely originated in the MW disc as supported by the small scale height of their vertical distribution and a substantial rotational velocity \citep[][]{Zinn.1985}.

The existence of a significant population of disc GCs gives us a chance to pin down the epoch of formation of the MW disc.
When reliable cluster ages have become available, a number of studies used GC distribution in the age-metallicity plane and their chemical element ratios to identify in-situ and accreted sub-populations \citep[][]{Marin_Franch.etal.2009,Forbes2010,Leaman2013,Recio-Blanco2018}. In particular, these studies identified a sequence of older in-situ GCs with disc-like kinematics and a sequence of younger clusters with steeper age-metallicity relation that was argued to be accreted with other galaxies. 

Some studies used available kinematic information to aid in-situ/accreted classification \citep[e.g.,][]{Dinescu.etal.1999}, but such information was quite limited until the advent of the Gaia satellite \citep{Gaia}. \citet{Massari2019} used distributions of GCs in the age-metallicity plane as guidance to come up with a number of criteria that use spatial distribution and kinematical properties of GCs measured by Gaia to classify almost all of the MW GCs into in-situ and accreted subpopulations. Some of the criteria used traditional cuts used in previous studies, such as a cut on the maximum $Z$ coordinate to identify ``disc'' clusters. Although reasonable, such criteria left a significant fraction of clusters with ambiguous/uncertain classification and these clusters were putatively assigned to new accreted structures (e.g., the ``low-energy group'') or known dwarf galaxies or streams. Some of these GCs were also argued to be a remnant of the putative massive dwarf galaxy that merged with the MW progenitor around $z>2$ \citep{Kruijssen.etal.2019,Kruijssen2020}. Thus, similarly to the early studies, in the recent classification attempts, a large fraction of the low-metallicity GCs with non-disc kinematics has been attributed to the accreted halo.

Recently, it was realized that in-situ born stars and stars in dwarf galaxies are distinct in their distributions of the aluminum-to-iron, $\rm [Al/Fe]$, and sodium-to-iron ratios \citep{Hawkins2015,Das2020}. We used this finding in \citet{Aurora} and \citet{Belokurov.Kravtsov.2023} to identify and study kinematic and chemical properties of the MW's in-situ stellar and GCs populations. The latter study showed that $\rm [Al/Fe]$-based classification at intermediate metallicites results in a fairly distinct distribution of the in-situ and accreted stars and GCs in the space of total energy $E$ and $L_z$ angular momentum and this can be used in the in-situ/accreted classification of the entire stellar and GC populations. 

In this study we use the $\rm [Al/Fe]$-calibrated in-situ/accreted classification in the $E-L_z$ plane to demonstrate that such classification results in the GC populations with distinct spatial, kinematic and chemical abundance distributions. We also show that the in-situ GCs in this classification show a clear disc spin-up signature that signals MW's disc formation and which was previously identified in the in-situ stars. 

The paper is organized as follows. In Section~\ref{sec:data} we describe the sample of GCs and their properties assembled from different sources. In Section~\ref{sec:class} we summarize the in-situ/classification method of \citet{Belokurov.Kravtsov.2023} and its underpinnings. We present distributions and statistics of the classified in-situ and accreted GC populations in Section~\ref{sec:results} and discuss differences from previous classification schemes in Section~\ref{sec:discussion}. We summarize our results and conclusions in Section~\ref{sec:conc}.  Finally, in the Appendix~\ref{app:chem_ref} we describe the data on chemical abundances from the literature that was used to complement APOGEE measurements. The Appendix~\ref{app:fire_facc_elz} presents results of the FIRE-2 simulations demonstrating the boundary between accretion and in-situ dominated regions in the energy-angular momentum plane. Appendix~\ref{app:spinup_fit} describes an alternative way to measure the spinup metallicity of GCs using a functional fit to the invididual $V_\phi$ and $\rm [Fe/H]$ of clusters. Finally, Appendix~\ref{app:class_tab} presents the table of the MW GCs with in-situ/accreted classifications according to the method presented in this paper.

\section{Sample of the Milky Way globular clusters}
\label{sec:data}


Our globular cluster catalogue is based on the 4th version of the GC database assembled by Holger Baumgardt. More specifically, we use i) the table with masses and structural parameters \footnote{See \url{https://people.smp.uq.edu.au/HolgerBaumgardt/globular/parameter.html}} and ii) the table with GC kinematics and orbital parameters \footnote{See \url{https://people.smp.uq.edu.au/HolgerBaumgardt/globular/orbits_table.txt}}, the latter table is used not only for the GCs' phase-space coordinates but also for the orbital eccentricities (computed with the published peri-centric and apo-centric distances). The resulting catalogue, once the two tables are cross-matched and merged, contains 165 Milky Way GCs. 

Mean cluster motions are based on Gaia EDR3 \citep[][]{gaia_edr3,Lindegren2021} data \citep[see][for details on the analysis of the Gaia EDR3 data]{Vasiliev2021}. This version uses the V-band luminosities derived in \citet{Baumgardt2020} and the GC distances derived in \citet{Baumgardt_Vasiliev2021}. Details of $N$-body models are described in \citet{Baumgardt2017} and \citet{Baumgardt2018}. Details on the stellar mass functions can be found in \citet{Baumgardt2023}. 

The catalogue is augmented with metallicities published by \citet{Harris2010} and other literature sources. GC ages used in this study are from \citet{VdB2013}. Total energy and the vertical component of the angular momentum $L_z$ for individual clusters used in the in-situ/accreted classification below are computed using the assumptions about the Galaxy  described in the beginning of Section~\ref{sec:data} of \citet{wrinkles}. The GCs in our catalogues are matched by name to 160 objects with tentative progenitor hosts published by \citet{Massari2019}. Five objects published by \citet{Massari2019}, namely Koposov 1, Koposov 2, BH 176, GLIMPSE 1 and GLIMPSE 2 are likely not of GC origin and therefore were not included in the GC database to begin with. Therefore only 155 out of 165 GCs in our catalogue have progenitor assignments in \citet{Massari2019}.

\section{Classification of accreted and in-situ  stars and clusters}
\label{sec:class}

Our method of classification of GCs and MW stars into in-situ and accreted clusters was presented in \citet{Belokurov.Kravtsov.2023}. Here we review the key details of the method relevant for this study. 

The method is based on classification of stars and clusters using the $\rm [Al/Fe]$ ratio. 
\citet{Hawkins2015} showed that $\rm [Al/Fe]$  have very different typical values in dwarf galaxies and in the Milky Way and argued that this difference can be used to distinguish the accreted and in-situ halo components \citep[see also][]{Das2020}. The difference arises because 
Al yield has a strong metallicity dependence and MW progenitor and dwarf galaxies that merge with it and contribute stars to the accreted halo component evolve at very different rates. The MW progenitor evolves fast and reaches metallicities required for efficient Al production much earlier than dwarf galaxies that form their stellar population at a much slower pace. As a result, stars born in the Milky Way exhibit a rapid increase in $\rm [Al/Fe]$ around $-1.5\rm <[Fe/H]<-0.9$ and there is a gap in the Al abundance at the same $\rm [Fe/H]$ between MW progenitor and dwarf galaxies that merge with it.

We use this fact as the basis for our classfication. 
Specifically, BK22 classify stars with $\rm [Al/Fe]>-0.075$ as in-situ and those with $\rm [Al/Fe]<-0.075$ as accreted, which is supported by the fact that the observed surviving massive MW dwarf satellites typically have $\rm [Al/Fe]<-0.1$ \citep{Hasselquist2021}. At metallicities $\rm [Fe/H]\lesssim-1.5$ the difference in typical values of $\rm [Al/Fe]$ between MW progenitor and accreted dwarfs becomes small and a clear $\rm [Al/Fe]$-based classification becomes unreliable. 
Our classification is thus based on a two-step approach. 

In the first step, we use the $\rm [Al/Fe]$-based classification in the metallicity range $-1.4<{\rm [Fe/H]}<-1.1$, where it is most reliable, with the threshold separating in-situ and accreted clusters assumed to be $\rm [Al/Fe]=-0.075$. As shown in Figure~2 of \citet[][see also discussion in their Section 3.2]{Belokurov.Kravtsov.2023} the in-situ and accreted components classified in this way, separate quite well in the plane of total energy $E$ and the $z$-component of the angular momentum $L_z$. This separation can be well described by the following $L_z$-dependent boundary in energy: 
\begin{align}
\label{eq:sel}
      L_z<-0.58:~ &	E=-1.3 \nonumber \\
-0.58<L_z<0.58:~ &	E=-1.4+0.3L_z^2\\
          L_z>0.58:~	&	E=-1.325+0.075L_z^2, \nonumber
\end{align}
where $E$ is in units of $10^5\, \mathrm{km}^2\,\mathrm{s}^{-2}$ and $L_z$ is in units of $10^3\,\mathrm{kpc}\,\mathrm{km}\mathrm{ s}^{-1}$. 

It is worth noting that although the form of this boundary is derived as an accurate empirical approximation to the boundary between regions of the $E-L_z$ space dominated by the in-situ and accreted populations in the $\rm [Al/Fe]$-based classification, a qualitatively similar boundary shape among the regions of the $E-L_z$ space that is dominated by these components is found in the FIRE-2 simulations of MW-sized galaxies (see Appendix~\ref{app:fire_facc_elz}).   

In \citet{Belokurov.Kravtsov.2023} we showed that this boundary does a good job of separating objects into in-situ and accreted components with a high accuracy ($\gtrsim 95\%$) comparable to the classification accuracy achievable with the best machine learning algorithms. Moreover, we have also tested that adding metallicity or Galactocentric distance to the classification with machine learning methods does not improve the accuracy. 

\begin{figure}
  \centering
  \includegraphics[width=0.5\textwidth]{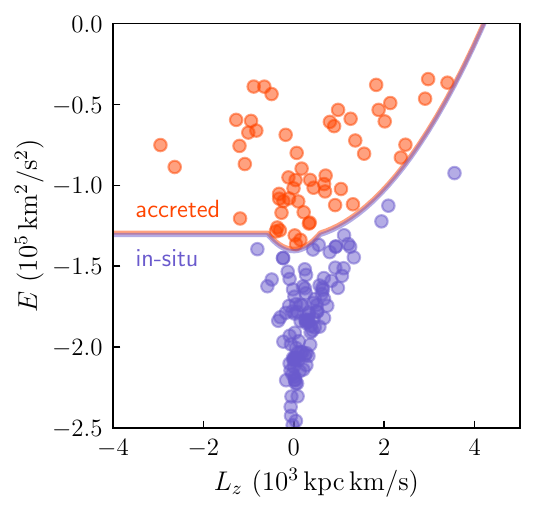}
  \caption[]{Distribution of the MW GCs in the plane of total energy $E$ and angular momentum $L_z$. The line indicates the boundary separating the in-situ clusters (blue) below the line and accreted clusters (red) above the line. }
   \label{fig:elz_class}
\end{figure}

Although this boundary is obtained using $\rm [Al/Fe]$ ratio within a limited range of metallicities, in the second step we assume that the same boundary $E_{\rm bound}(L_z)$ is applicable in the entire metallicity range. 
Figure~\ref{fig:elz_class} shows $E$ and $L_z$ distributions of the MW GCs classified as accreted (primarily GS/E, red) and in-situ (blue) along with the boundary $E_{\rm bound}(L_z)$ used for classification.  In what follows, we examine distribution of various GC properties in the components classified using the boundary shown in Figure~\ref{fig:elz_class}.

Out of 164 Galactic GCs with measured metallicities considered in this study, 106, or $\approx 2/3$ are classified as in-situ and 58 as accreted. Classification for individual clusters is presented in the Table~\ref{tab:class} in Appendix~\ref{app:class_tab}.

\begin{figure*}
  \centering
  \includegraphics[width=0.99\textwidth]{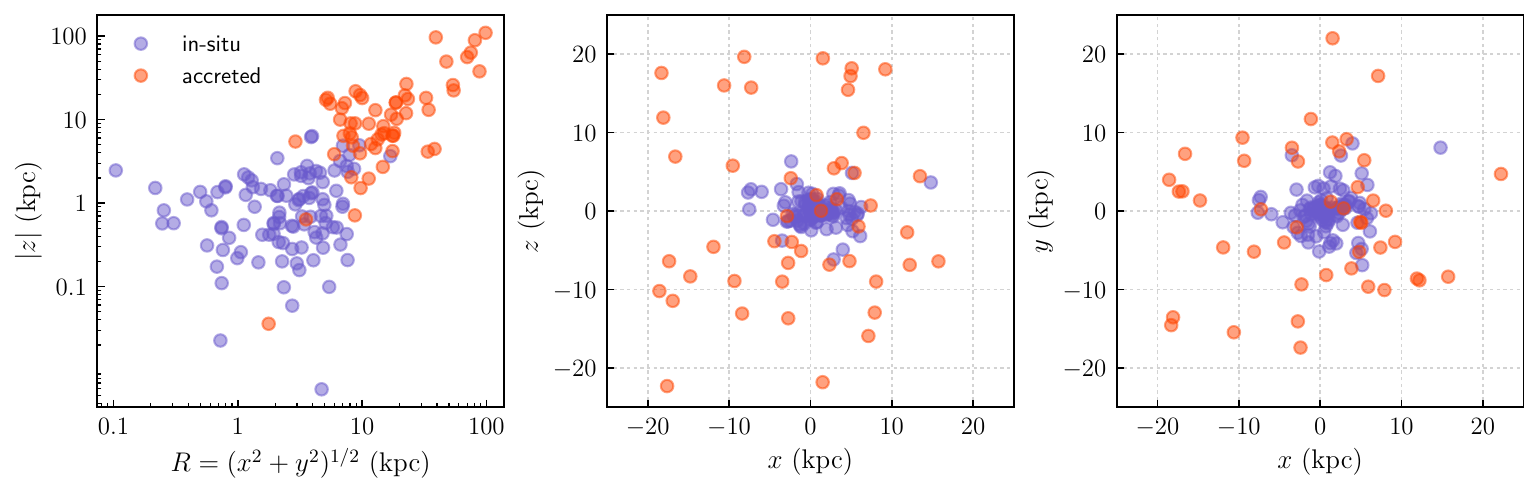}
  \caption[]{Spatial distribution of the in-situ (blue) and accreted (red) MW GCs classified using method described in Section~\ref{sec:class} (see Fig.~\ref{fig:elz_class}). The left panel shows the absolute value of $z$ coordinate (in the coordinate system where MW disc is in the $x-y$ plane) as a fraction of galactocentric distance in the disc plane $R=(x^2 +y^2)^{1/2}$. The middle and right panels show spatial distributions of the in-situ and accreted GCs in the $x-z$ and $x-y$ plane. The figure shows that the two populations have very distinct spatial distributions. Note that the distribution of the accreted clusters is fairly isotropic, while the distribution of in-situ clusters is flattened around the $x-y$ plane of the MW disc. }
   \label{fig:spatial}
\end{figure*}
%

\section{Results}
\label{sec:results}

In what follows we will consider distributions of various properties of the in-situ and accreted GCs in our classification from their spatial distributions to the distributions of their metallicity, age and kinematic properties. 

\subsection{Spatial distribution}
\label{sec:spatial}

Figure~\ref{fig:spatial} shows spatial distribution of the in-situ (blue) and accreted (red) MW GCs classified using method described above in Section~\ref{sec:class} (see Fig.~\ref{fig:elz_class}). The left panel shows the absolute value of $z$ coordinate (in the coordinate system where MW disc is in the $x-y$ plane) as a fraction of galactocentric distance in the disc plane $R=(x^2 +y^2)^{1/2}$. It shows clearly that the two populations are well segregated in both $z$ and $R$ with most of the in-situ classified clusters located at $\vert z\vert<3$ kpc and $R<10$ kpc. The distribution of the accreted clusters, on the other hand, is much more extended. 
Although one could argue that such segregation is largely defined by the fact that classification of clusters is done using total energy boundary, the fact that the in-situ clusters have different $\vert z\vert$ and $R$ ranges shows that it does not simply classify clusters within a given limiting galactocentric distance. 

Indeed, as can be seen in 
the middle panel of Figure~\ref{fig:spatial} the distribution of in-situ clusters is quite flattened in the $z$ direction around the $x-y$ plane of the MW disc.
The distribution of the accreted clusters, on the other hand, is fairly isotropic. 
The right panel does not show a similar flattening in the in-situ GC distribution in the $x-y$ projection indicating that it can be characterized as an oblate ellipsoid or a thick disc. As we will discuss below in Section~\ref{sec:lohicomp}, the discy flattened distribution is even more pronounced for the in-situ clusters with $\rm [Fe/H]>-1$. 

The figure shows that the two classified populations have very distinct spatial distributions. Notably, similar two populations are identified if we use the OPTICS clustering algorithm \citep{OPTICS} and reachability curves\footnote{For examples of application of the OPTICS algorithm see \url{https://scikit-learn.org/stable/auto_examples/cluster/plot_optics.html}} to identify clusters in the 3D distribution of GCs, which again indicates good segregation of the in-situ and accreted populations in space.

\subsection{Distribution in the age-[Fe/H] plane}
\label{sec:agefeh}

%
\begin{figure}
  \centering
  \includegraphics[width=0.5\textwidth]{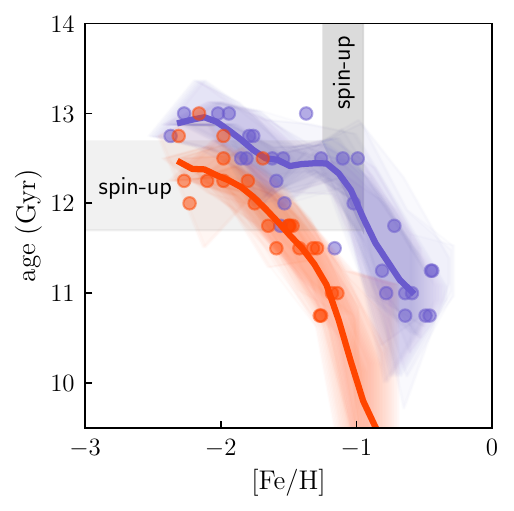}
  \caption[]{Distribution of MW GCs classified as in-situ (blue) and accreted (red) by our classification method that have age estimates by \citet{VdB2013}. The circles show the individual GCs, while the lines are the median of the binned distributions (see text for details). The shaded rectangular areas indicate the spin-up range of metallicities and the corresponding approximate range of the stellar age (i.e., lookback spin-up time). }
   \label{fig:agefeh}
\end{figure}

Figure~\ref{fig:agefeh} shows distributions of the in-situ (blue) and accreted (red) MW GCs in the age-metallicity plane using clusters that have age estimates \citep{VdB2013}. The circles show the individual GCs, while the lines are the median of the binned distributions obtained using different radial range for bin placement. Specifically, we shift all bin edges to the right in small increments from their original locations up to the shift equal to the bin size and reconstruct histograms for the new edges. We then estimate the median of all the histograms obtained for individual shifts and 68\% region of histograms around the median. 

The figure shows that our classification in the total energy and angular momentum $L_z$ selects distinct sequences of clusters in the age--metallicity plane with a rather small overlap. The in-situ clusters are predominantly older by $0.5$ Gyr at a given metallicity for $\rm [Fe/H]\lesssim -1.5$ and by $\gtrsim 1$ Gyr at larger metallicities. Conversely, the in-situ clusters have larger metallicities by $\approx 0.3-0.5$ dex at a given age. The two sequences overlap somewhat only for the oldest and lowest metallicity clusters.  

The two sequences in the age-metallicity space identified by our classification in the $E-L_z$ plane were identified previously \citep[see discussion in e.g.][]{Forbes2010}. Notably, \citet{Leaman2013}  identified a clear sequence of GCs born with disc-like kinematics down to $\rm [Fe/H]\approx -1.3$ \citep[see also][]{Recio-Blanco2018}. As we discuss 
below, the kinematics of these GCs is consistent with their disc origin. Our classification shows that these clusters are a part of the ``in-situ sequence'' that extends to metallicities of $\rm [Fe/H]\approx -2.3$. This is consistent with the model-based interpretation of \citet{Kruijssen.etal.2019b}.

It is worth noting that the in-situ sequence in Figure~\ref{fig:agefeh} has a sharp turnover to lower ages at $\rm [Fe/H]\approx -1$. Although this metallicity is similar to the metallicity of the disc spin-up that we will discuss below, this turnover is likely not directly related to disc formation but reflects the general form of the age-metallicity relation of MW-sized galaxies. Indeed, galaxy formation models generally predict such turnover exacty at $\rm [Fe/H]\approx -1$ (see, e.g., the middle panel of Figure 14 in BK22), which marks the transition from the fast to slow mass accretion regime of evolution. 

The gray-shaded vertical rectangular area in the Figure shows the range of metallicities $\rm [Fe/H]\approx -1.3\div -0.9$ which corresponds to the disc spin-up exhibited by the in-situ MW stars estimated in BK22. The horizontal gray rectangular area shows the corresponding approximate range of cluster ages of $\approx 11.7-12.7$ Gyr (i.e., lookback spin-up time) consistent with the estimate of \citet{Conroy2022}. This range of disc formation lookback times corresponds to the range of redshifts $z\approx 3.07-5.3$ for the Planck cosmology. Although this range is fairly broad, the result indicates that Milky Way formed its disc earlier than a typical galaxy of similar stellar mass both in observations \citep{Simons.etal.2017}
and galaxy formation simulations \citep[][]{Aurora,Semenov.etal.2023a,Dillamore.etal.2023b}.

\subsection{Metallicity distributions of in-situ and accreted GCs}
\label{sec:fehdist}

\begin{figure}
  \centering
  \includegraphics[width=0.5\textwidth]{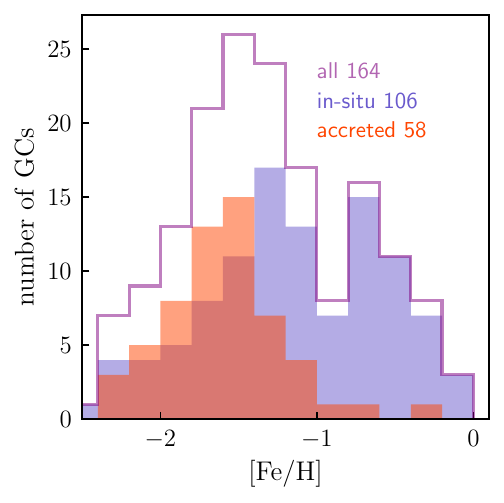}
  \caption[]{Distribution of $[\rm Fe/H]$ of the MW GCs (solid magenta histogram), in-situ GCs (blue histogram) and accreted GCs (red histogram) in our classification. Note that bimodality in the metallicity distribution is only present in the in-situ GCs. }
   \label{fig:feh_dist}
\end{figure}

Metallicity distributions of the in-situ and accreted GCs are shown in Figure~\ref{fig:feh_dist} along with the metallicity distribution of the entire GC sample. The distributions of the in-situ and accreted GCs in our classification are clearly different with accreted GCs having mostly metallicities $[\rm Fe/H]<-1$. At the same time, at these low metallicities there is a significant overlap
of the accreted and in-situ clusters and they clearly do not separate neatly in metallicity, as envisioned in the classification of \citet{Zinn.1985}.

Remarkably, Figure~\ref{fig:feh_dist} shows that only the distribution of in-situ GC metallicities is bimodal, while accreted clusters have a distribution with a single peak at   $[\rm Fe/H]<-1.6$. 
The origin of the bi-modality in the metallicity distribution is still debated. However, it may be imprinted in the $[\rm Fe/H]$ distribution by the same transition from the fast to slow mass accretion regime that produces the sharp turnover in the age-$[\rm Fe/H]$ sequence of the in-situ clusters discussed above. After the Galaxy transitions to the slow accretion regime at  $[\rm Fe/H]\approx -1$ clusters born with a broad range of ages have similar metallicities, which creates a peak at high metallicities \citep[see][]{El_Badry_etal.2019}. Conversely, clusters born during the early fast accretion regime have a narrow range of ages and broad distribution of metallicities, which peaks at metallicities corresponding to the time when the MW progenitor's star formation rate was at its maximum.   

\begin{figure}
  \centering
  \includegraphics[width=0.5\textwidth]{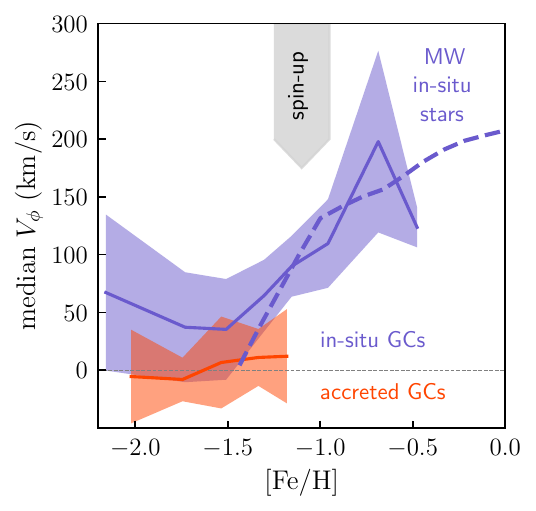}
  \caption[]{Tangential velocity of the in-situ (blue) and accreted (red) GCs as a function of their metallicity $\rm [Fe/H]$. The solid lines show the median of the bootstrap resamples of the original GC sample, while shaded areas show their $1\sigma$ scatter. The dashed line shows the corresponding median $V_\phi$ as a function of metallicity for the in-situ MW stars, as estimated by \citep{Aurora}. Downward gray arrow indicates the range of metallicities in which MW disc spin-up was identified for these stars. The figure shows that the in-situ MW GCs in our classification also exhibit a clear spin-up feature at the same metallicity range of $\rm [Fe/H]\in[-1.3,-0.9]$ as the in-situ stars.  }
   \label{fig:vphifeh}
\end{figure}
\begin{figure}
  \centering
  \includegraphics[width=0.5\textwidth]{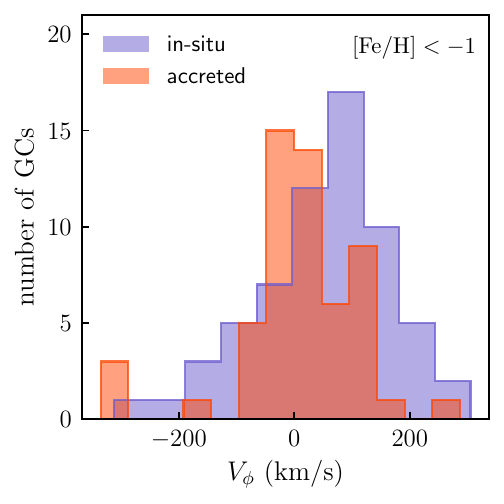}
  \includegraphics[width=0.5\textwidth]{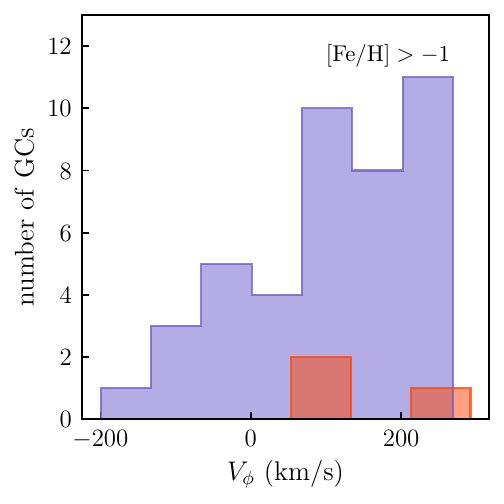}
  \caption[]{Distributions of the tangential velocity $V_\phi$ for the in-situ (blue) and accreted (red) MW GCs. The upper panel shows distribution for the low-metallicity GCs with $\rm [Fe/H]<-1$, while the lower panel shows the distribution for GCs with $\rm [Fe/H]>-1$. The figure shows that $V_\phi$ distributions of low- and high-metallicity clusters are quite different. Distribution for the low-metallicity clusters has a single peak, with that of the accreted clusters centered at $V_\phi\approx 0$ km/s, while distribution for the in-situ clusters centered at $V_\phi\approx 50$ km/s. The $V_\phi$ distribution of the high-metallicity in-situ clusters is very skewed with a significant fraction of clusters coherently rotating with $V_\phi\sim 200$ km/s, while a tail of the in-situ GCs has $V_\phi\lesssim 0$. }
   \label{fig:vphidist_inex}
\end{figure}
%

%
%

%
\subsection{Milky Way disc spin-up traced by in-situ GCs}
\label{sec:gcspinup}

Figure~\ref{fig:vphifeh} shows the tangential velocity of the in-situ (blue) and accreted (red) GCs as a function of their metallicity $\rm [Fe/H]$. The solid lines show the median of the bootstrap resamples of the original GC sample, while shaded areas show their $1\sigma$ scatter. The dashed line shows the corresponding median $V_\phi$ as a function of metallicity for the in-situ MW stars, as estimated by \citet{Aurora}. Although the number of clusters per bin is fairly small and exact form of the median $V_\phi$ curve depends on the number of bins used, in the Appendix \ref{app:spinup_fit} we show that a similar result is obtained if a parametric ``soft step'' function is fit to the distribution of individual $\rm [Fe/H]$ and $V_\phi$ values. 

The figure shows that the in-situ MW GCs in our classification also exhibit a clear spin-up feature at the same metallicity range of $\rm [Fe/H]\in[-1.3,-0.9]$ as the in-situ stars of the Milky Way. The fact that metal-rich ``disc'' GCs ($\rm [Fe/H]>-0.8$ in the \citealt{Zinn.1985} classification) exhibit large net rotation is well-known \citep{Armandroff.1989}. Figure~\ref{fig:vphifeh}, however, shows that the process of the MW disc formation is imprinted in its in-situ GC population at metallicities $\rm [Fe/H]\gtrsim -1.3$. 
This implies the in-situ GCs at this wide range of metallicities were formed in the MW disc after its formation and retained corresponding kinematics \citep[see also][]{Leaman2013,Recio-Blanco2018}. 

Conversely, the in-situ GCs with metallicities $\rm [Fe/H]<-1.3$ were born during turbulent pre-disc stages of MW evolution and are thus a part of the {\it Aurora} stellar component of the Galaxy identified in \citet[][see also \citealt{Conroy2022,Rix2022}]{Aurora}.\footnote{Named after Aurora -- the Latin name of the goddess of dawn Eos in Greek mythology.} Interestingly, the Aurora clusters show net rotation with the median $V_\phi\approx 50$ km/s.  This net velocity is similar to the typical median velocity of in-situ at the pre-disc metallicites in simulations of MW-sized galaxies \citep[][]{Aurora, Semenov.etal.2023a,Dillamore.etal.2023b}. However, the non-zero median $V_\phi$ does not imply that these stars and GCs were born in a disc.  In fact, they were generally born in very chaotic configurations \citep[see Fig. 10 in][]{Aurora}. Nor does it necessarily mean that GCs were born with such net rotation. 
As shown by \citet{Dillamore.etal.2023}, its origin maybe in the trapping of these old GCs by rotating bar that forms during latter stages of the MW disc evolution. 

\begin{figure}
  \centering
  \includegraphics[width=0.5\textwidth]{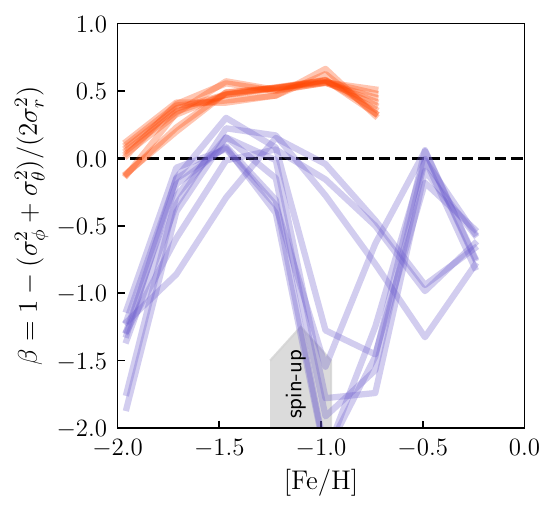}
  \caption[]{Velocity anisotropy of the MW GCs defined as $\beta=1-(\sigma^2_\phi+\sigma^2_\theta)/(2\sigma^2_r)$ as a function of metallicity $\rm [Fe/H]$ for the accreted (red) and in-situ (blue) GCs. Different lines correspond to the estimates obtained using different placement of metallicity bins in the range of metallicities spanned by the GCs. }
   \label{fig:betafeh}
\end{figure}

Figure~\ref{fig:vphidist_inex} shows distributions of the tangential velocity $V_\phi$ for the in-situ (blue) and accreted (red) MW GCs. The upper panel shows distribution for the low-metallicity GCs with $\rm [Fe/H]<-1$, while the lower panel shows the distribution for GCs with $\rm [Fe/H]>-1$. The figure shows that $V_\phi$ distributions of low- and high-metallicity clusters are quite different. The distribution for the low-metallicity clusters has a single peak, with that of the accreted clusters centered at $V_\phi\approx 0$ km/s, while distribution for the in-situ clusters centered at $V_\phi\approx 50$ km/s as noted above. 

The $V_\phi$ distribution of the high-metallicity in-situ clusters is very skewed with a significant fraction of clusters coherently rotating with $V_\phi\sim 200$ km/s, while a tail of the in-situ GCs has $V_\phi\lesssim 0$. Note that $V_\phi$ distributions of the low- and high-metallicity in-situ GCs is very similar to the distribution of tangential velocities of the MW's in-situ stars in Figure~6 of \citet{Aurora} at similar metallicities. In particular, in-situ stars also exhibit a tail towards $V_\phi<0$ and a similar tail can be seen in the distribution of in-situ stars in simulations of the MW-sized galaxies (see Appendix~\ref{app:fire_facc_elz}). 


%
\begin{figure*}
  \centering
  \includegraphics[width=\textwidth]{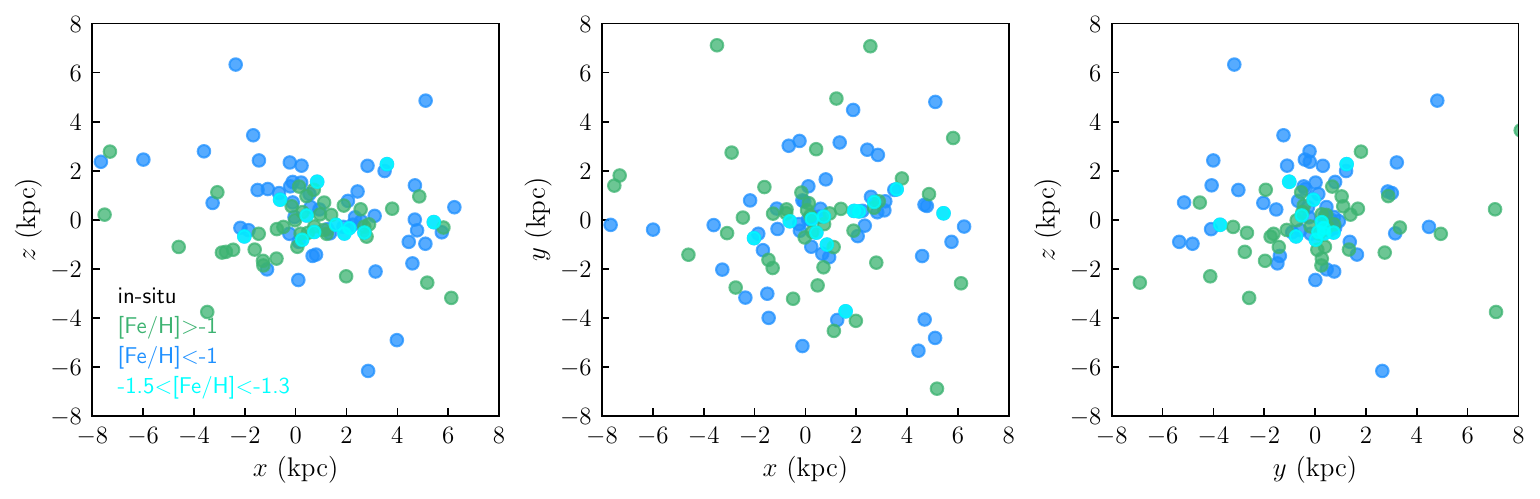}
  \caption[]{Projections of the spatial distribution of GCs classified as in-situ in different ranges of metallicity: {\it blue points} show clusters with $\rm [Fe/H]<-1$, of these clusters in the metallicity range of the first peak in the metallicity distribution, $-1.5<\rm [Fe/H]<-1.3$, shown by {\it points of cyan color}, {\it green points} show GCs with $\rm [Fe/H]>-1$. The figure shows that the distribution of high-metallicity in-situ GCs is more flattened around the $z=0$ plane than the distribution of low-metallicity clusters. It also shows that 11 GCs in the metallicity range $-1.5<\rm [Fe/H]<-1.3$ are distributed in a rather narrow filament or prolate ellipsoid with a small $c/a$ axes ratio. }
   \label{fig:spatial_in_lohi}
\end{figure*}

Finally, Figure~\ref{fig:betafeh} shows velocity anisotropy  defined as 
\begin{equation}
\beta=1-\frac{\sigma^2_\phi+\sigma^2_\theta}{2\sigma^2_r}     
\end{equation}
as a function of metallicity $\rm [Fe/H]$ for the accreted (red) and in-situ (blue) GCs. Different lines correspond to the estimates obtained using different placements of the metallicity bins in the range spanned by the GCs. It shows that velocity anisotropy of the accreted clusters is close to isotropic at the lowest metallicity and has a moderate radial anisotropy
at metallicities $\rm [Fe/H]\approx -1.7\div-0.7$. The in-situ GCs, on the other hand, have a nearly isotropic velocity distribution at $\rm [Fe/H]\lesssim -1.3$, but the distribution changes sharply at lower metallicities where the distribution has a clear tangential anisotropy.

\subsection{Comparisons of the low- and high-metallicity in-situ clusters}
\label{sec:lohicomp}

Given qualitative changes that MW progenitor clearly underwent at $\rm [Fe/H]=-1$ both due to the transition from the fast to slow mass accretion regime and due to the formation of the disc, it is interesting to consider differences in properties of the in-situ GCs with $\rm [Fe/H]<-1$ and $\rm [Fe/H]>-1$ straddling this transition metallicity. 

Figure~\ref{fig:spatial_in_lohi} shows the $x-z$, $x-y$, and $y-z$ projections of the spatial distribution of in-situ GCs with metallicities $\rm [Fe/H]<-1$ and $\rm [Fe/H]>-1$. The figure shows that the distribution of high-metallicity in-situ GCs is somewhat more flattened around the $z=0$ plane than the distribution of low-metallicity clusters consistent with their formation in the disc.

Interestingly, we also find that 11 clusters in the metallicity range of the first peak in the metallicity distribution $-1.5<\rm [Fe/H]<-1.3$ (shows as cyan points) are distributed in a rather narrow filament or prolate ellipsoid with a small $c/a$ axes ratio. Although the number of objects is too small to make definitive conclusions, we speculate that the formation of these clusters could have been induced in the MW progenitor by the tidal forces and/or gas accretion associated with the early stages of the GS/E merger. This process thus could be responsible for both an overall burst of 
star formation in the MW progenitor and burst of GC  formation that produced the low-metallicity peak in the metallicity distribution of in-situ clusters (see Fig.~\ref{fig:feh_dist}). Indeed, one can generally expect that the maximum initial mass of the forming GCs scales with star formation rate \citep{Maschberger.Kroupa.2007} and initial in-situ GC masses estimated by \citet{Baumgardt2003} at metallicities $-1.5\lesssim \rm [Fe/H]\lesssim -1.3$ do reach values of $\log_{10} M_{\rm ini}>6.5$ larger than maximum initial masses for neighboring metallicity ranges. In fact, all of the other MW GCs with such large masses are among in-situ clusters at $\rm [Fe/H]>-1$ near the second peak in their metallicity distribution. It is also notable that accreted GCs do not show such increased maximum $M_{\rm ini}$ at any metallicity. 

\begin{figure*}
  \centering
  \includegraphics[width=\textwidth]{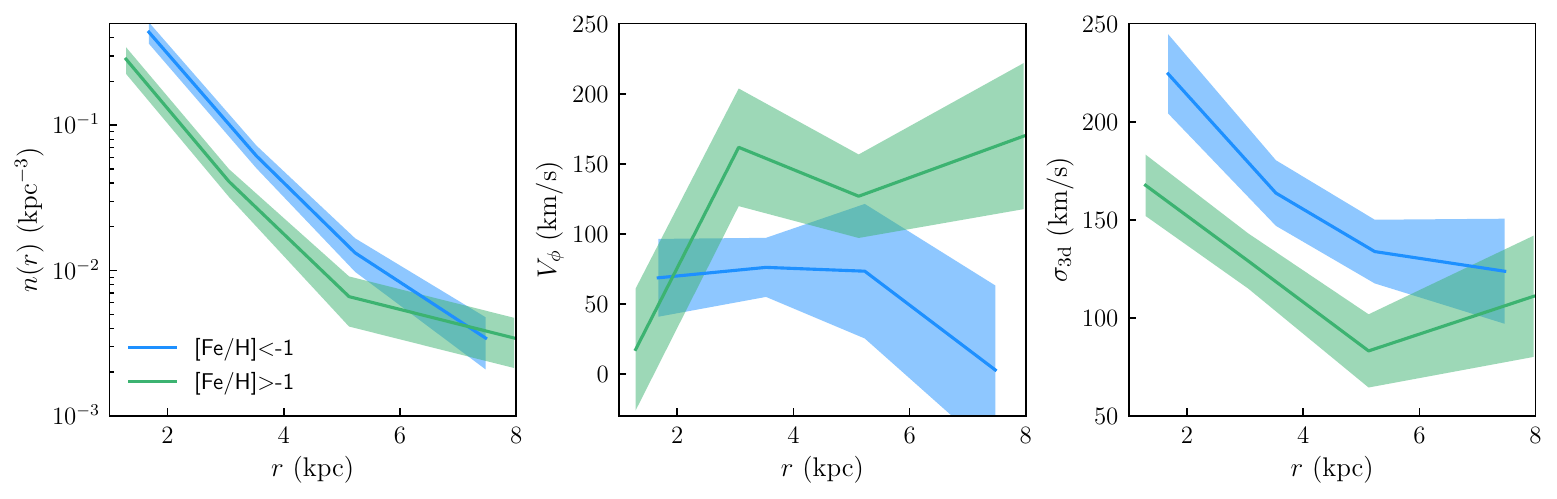}
  \caption[]{Profiles of number density (left panel), median tangential velocity (middle panel) and 3d velocity dispersion (right panels) of the in-situ GCs with metallicities $\rm [Fe/H]<-1$ (blue) and $\rm [Fe/H]<-1$ (green). The lines show the median profiles of bootstrap samples while shaded regions show standard deviation of the median profiles of the bootstrap samples. The left panel shows that radial distribution of the high-metallicity in-situ clusters is more concentrated, while the right panel shows that velocity dispersion of the high-metallicty GCs is considerably lower than that of the low-metallicity in-situ clusters. The middle panel shows that high-metallicity in-situ GCs population exhibits coherent rotation with $V_\phi(r)$ reminiscent of a rotation curve, while low metallicity in-situ GCs also show coherent net rotation with  a much smaller value of $\approx 50$ km/s. }
   \label{fig:hilo_pro}
\end{figure*}

Figure~\ref{fig:hilo_pro} shows comparisons of the radial number density profiles, tangential velocity, and 3d velocity dispersion profiles of the in-situ GCs with metallicities $\rm [Fe/H]<-1$ (blue) and $\rm [Fe/H]<-1$ (green). The lines show the median profiles of bootstrap samples while shaded regions show the standard deviation of the median profiles of the bootstrap samples. 

The figure shows that the radial distribution of the high-metallicity in-situ clusters is more concentrated, while their velocity dispersion is considerably lower than that of the low-metallicity in-situ clusters. This is because higher metallicity clusters formed within a relatively compact MW disc, while $\rm [Fe/H]<-1$ clusters formed during chaotic pre-disc stages of evolution and were likely dynamically heated both by mergers and by feedback-driven inflows and outflows. As we noted above, they were also likely affected by the Milky Way bar which induced a small net $V_\phi\approx 50$ km/s velocity \citep[see][]{Dillamore.etal.2023}.

Likewise, the $V_\phi(r)$ profile comparisons in the middle panel shows that high-metallicity in-situ GCs population exhibits coherent rotation with $V_\phi(r)$ reminiscent of a rotation curve, while low metallicity in-situ GCs also show coherent net rotation but with a much smaller value of $\approx 50$ km/s, in agreement with the change of $V_\phi$ as a function of $\rm [Fe/H]$ in Figure~\ref{fig:vphifeh}.

%

\section{Discussion}
\label{sec:discussion}

\subsection{Comparison with previous classifications}
\label{sec:prevcomp}

%
\begin{figure}
  \centering
  \includegraphics[width=0.5\textwidth]{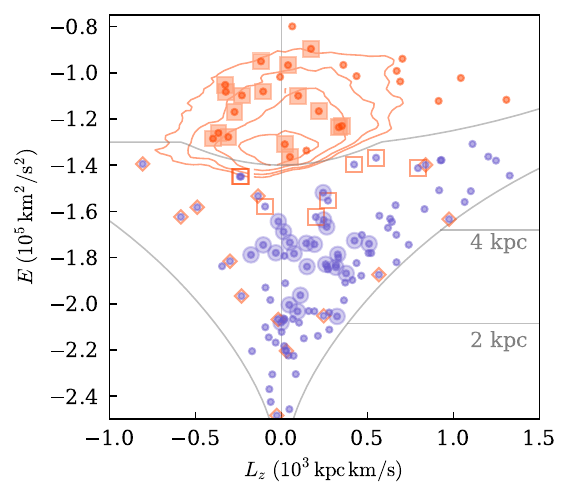}
  \caption[]{Zoom-in on the low-$E$ portion of the $E-L_z$ plane. Small blue (orange) filled circles mark the locations of in-situ (accreted) GCs according to our classification. Filled (empty) squares are accreted (in-situ) clusters assigned to the GS/E merger by \citet{Massari2019}. Large filled circles mark those GCs that are classified as in-situ in our method but as the `low-energy group' by \citet{Massari2019}. Diamonds are in-situ clusters without a well-defined classification in \citet{Massari2019}. Grey curves give maximal $L_z$ at a given $E$. Orange contours give the location of the GS/E tidal debris isolated in the {\it Gaia} DR3 data by \citet{wrinkles}.
}
   \label{fig:elz_zoom}
\end{figure}

As noted in Section~\ref{sec:intro}, a number of previous studies devised methods to classify accreted and in-situ GCs using properties of GCs. The study most relevant for comparison with our classification method is \citet{Massari2019} because it uses similar cluster properties for classification and we thus focus on the comparison with their classification here. 

All of the GCs we classify as accreted (58 in total) are also classified as accreted by \citet{Massari2019}. However, this study classifies only 61 GCs among our in-situ sample as in-situ. The rest is classified as accreted or undetermined. Below we focus on these low-energy objects and discuss observational clues to their origin.

Figure~\ref{fig:elz_zoom} zooms in on the portion of the $E-L_z$ space just below the in-situ/accreted decision boundary where we indicate the assignment adopted by \citet{Massari2019}. Only 61 of 107 classified as in-situ in our method (small blue-filled circles) formed in the MW according to \citet{Massari2019}: they classified 25 of these clusters as ``the disc'' (M-D, following their designation) and 36 as ``the bulge'' (M-B). 

The other 46 in-situ clusters in our classification are classified by \citet{Massari2019} as follows. 24 clusters (large blue circles) are assigned to the ``low-energy group'', which was later interpreted to be a signature of an accretion event at $z\sim 1$, sometimes referred to as Kraken \citep[][]{Kruijssen.etal.2019,Kruijssen2020} or Koala \citep[][]{Forbes2020}. There are also 8 GCs assigned by \citet{Massari2019} to the GS/E merger, but classified as in-situ in our scheme; these are marked with empty orange squares. Finally, there are 14 GCs with undetermined classification in \citet{Massari2019}, marked with orange diamonds: these either have '?' or 'XXX' for the possible progenitor or were not included in their catalogue (7 out of 14). 

\begin{figure*}
  \centering
  \includegraphics[width=\textwidth]{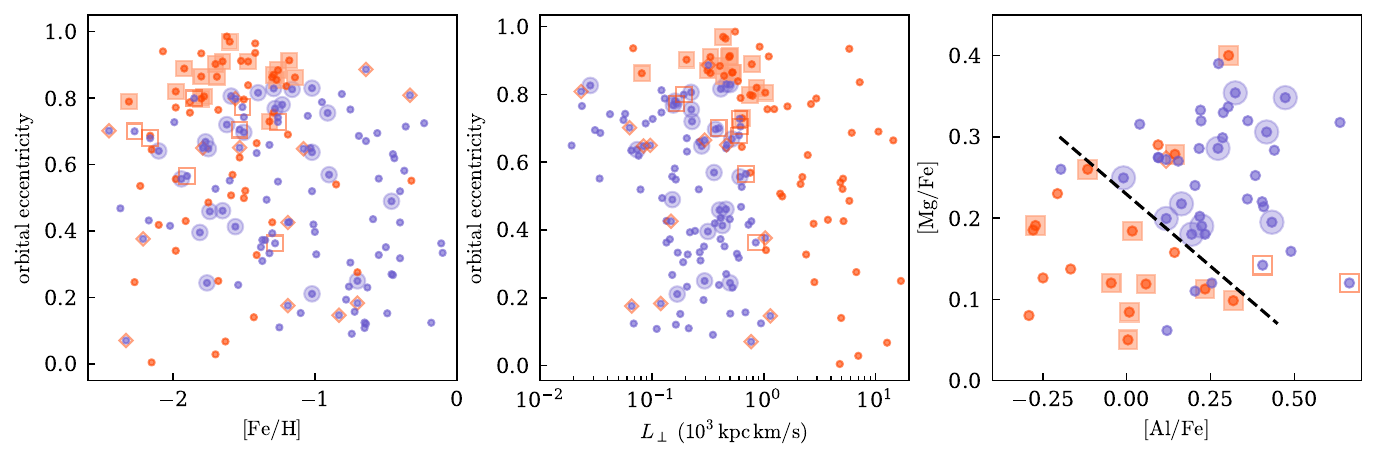}
  \caption[]{Orbital and chemical properties of in-situ and accreted GCs. {\bf Left:} Orbital eccentricity as a function of metallicity.  {\bf Middle:} Eccentricity as a function of $L_{\perp}$ component of the angular momentum. {\bf Right:} [Mg/Fe] vs [Al/Fe] for GCs with available abundance measurements (see Appendix~\ref{app:chem_ref} for comparison between APOGEE and literature values).  Small blue (orange) filled circles mark the locations of in-situ (accreted) GCs according to our classification. Filled (empty) squares are accreted (in-situ) clusters assigned to the GS/E merger by \citet{Massari2019}. Large filled circles mark those GCs that are classified as in-situ in our method but as the `low-energy group' by \citet{Massari2019}. Diamonds are in-situ clusters without a well-defined classification in \citet{Massari2019}.
  }
   \label{fig:ecc_mg_al}
\end{figure*}

To gain a better perspective on the chemo-dynamic properties of these various GC groups,  Figure~\ref{fig:ecc_mg_al} shows cluster orbital eccentricity (computed with the pericentre and apocentre estimates from the H. Baumgardt database, see Section~\ref{sec:data}) as a function of metallicity in the left panel, eccentricity as a function of $L_{\perp}=\sqrt{L_x^2+L_y^2}$ -- the component of the angular momentum perpendicular to $L_z$ -- in the middle panel, and the distribution of the GCs (where abundance measurement is available) in the plane of [Mg/Fe] and [Al/Fe] in the right panel. Here we have included several literature values that were corrected to the APOGEE abundance scale, as described in Appendix~\ref{app:chem_ref}. The GCs with non-APOGEE measurements of [Al/Fe] and [Mg/Fe] from the literature include NGC 1261 \citep{ngc1261}, Rup 106 \citep{rup106}, NGC 4833 \citep{ngc4833}, NGC 5286 \citep{ngc5286}, NGC 5466 \citep{ngc5466}, NGC 5927 \citep{ngc5927}, NGC 5986 \citep{ngc5986}, NGC 6139 \citep{ngc6139}, NGC 6229 \citep{ngc6229}, NGC 6266 \citep{ngc6266}, NGC 6355 \citep{ngc6355}, NGC 6362 \citep{ngc6362}, NGC 6402 \citep{ngc6402}, NGC 6440 \citep{ngc6440}, NGC 6522 \citep{ngc6522}, NGC 6528 \citep{ngc6528}, NGC 6584 \citep{ngc6584}, NGC 6624 \citep{ngc6624}, NGC 6864 \citep{ngc6864} and NGC 6934 \citep{ngc1261}.

Recently, we showed that in-situ and accreted GCs separate well in the space of [Mg/Fe]-[Al/Fe] \citep{Belokurov.Kravtsov.2023}. In addition, at [Fe/H]$>-2$ the in-situ stars have higher values of [Mg/Fe] compared to those accreted \citep{Aurora}. This trend,  however, is blurred by the internal GC evolution where Mg can be destroyed to make Al. As a result, clusters may end up having lower values of [Mg/Fe]. Nevertheless, the anomalous chemistry is betrayed by their elevated [Al/Fe] ratio. Thus, in the plane of [Mg/Fe]-[Al/Fe] GCs can move diagonally from top left to bottom right, as indicated by the black dashed line. While the chemical plane of [Al/Fe] and [Mg/Fe] appears to work well to separate the GCs into two distinct groups, it would be beneficial to explore the use of elemental abundances not affected by the cluster's secular evolution. In connection to this, most recently, other chemical tags have been proposed to pin down the origin of the Galactic GCs. For example, \citet{Minelli2021} advocate the use of Sc, V, and Zn, while Monty et al (2023, in prep) show that Eu can be used as a strong tag of the GS/E GCs.

Only two GCs classified as GS/E in our scheme (NGC 288 and NGC 5286) lie in the top right corner of the [Mg/Fe]-[Al/Fe] plane shown in the right panel of Figure~\ref{fig:ecc_mg_al}. For both of these clusters recent chemical abundance measurements indicate that these clusters are probably not associated with the GS/E merger \citep[see][]{Monty2023}. There is an additional cluster, NGC 6584, which lacks a clear progenitor in the classification of \citet{Massari2019} and is classified as accreted in our scheme. 

Four GCs classified as in-situ in our scheme using the $E-L_z$ boundary lie in the bottom left corner of the [Mg/Fe]-[Al/Fe] plane dominated by accreted clusters (although three are close to the nominal boundary). Assuming all four are indeed misclassified and were accreted, the fraction of accreted clusters among GCs classified as in-situ by our scheme can be classified as $\approx 4/39\approx 10\%$, where 39 is the number of in-situ GCs above the dashed line in the [Mg/Fe]-[Al/Fe] plane. 

Focusing on the ``low-energy group'' of globular clusters (large blue circles), it is difficult to see how these objects can be a part of a single accretion event. These 24 GCs do not cluster together in any of the orbital properties considered. Instead, they span a large range of E,  $L_z$, $L_{\perp}$, and eccentricity. This is in stark contrast with the GS/E highlighted with filled orange squares: these GCs have a narrow range of eccentricity and $L_{\perp}$. In terms of $L_z$, the low energy group GCs appear to have an extent similar to the GS/E members. This, however, is an illusion: the available range of $L_z$ is a strong function of energy and drops with decreasing $E$. For the energy level of the clusters labeled as the `low-energy group', the $L_z$ range is less than half of that at the level of the GS/E GCs. Therefore, the relative $L_z$ dispersion of these GCs is larger by more than a factor of 2. These clusters also have a clear net prograde motion with a mean $V_{\phi}\approx60$ km/s similar to the bulk of the in-situ GCs and typical of the Aurora population. 

In terms of their chemistry, all 10 (out of 24) of the `low-energy group' GCs with available abundance measurements lie with the rest of the in-situ clusters in the [Mg/Fe]-[Al/Fe] plane shown in the right panel of Figure~\ref{fig:ecc_mg_al}. We conclude that there is no strong evidence in favour of a distinct low-energy 
group of clusters because in every property considered, these clusters span the range typical of in-situ GCs. 

Note that the main reason these clusters were classified as accreted by \citet{Massari2019} is because they are located outside of the nominal ``bulge'' radius of 3.5 kpc. However, this adopted size is rather arbitrary because the peanut bulge of the Milky Way has a radial extent of $\approx 1.5$ kpc, while at larger distances stellar distribution is arranged into a prominent bar \citep[see, e.g., Fig. 1 in ][]{Wegg2015,Barbuy.etal.2018}. Incidentally, in our catalogue, 14 out of the 24 GCs assigned to the low-energy group by \citet{Massari2019} have Galactocentric distances smaller than 3.5 kpc (also see Figure~\ref{fig:elz_zoom}).

Let us now briefly consider the 8 GCs (highlighted with open orange squares) classified as in-situ in our scheme, but associated with the GS/E event by \citet{Massari2019}. As Figure~\ref{fig:elz_zoom} illustrates, the density of the GS/E stellar component drops abruptly below $E=-1.4\times10^5 {\rm km}^2/{\rm s}^2$. Most of the 8 alleged GS/E GCs lie outside of the orange contours and thus have values of total energy lower than the bulk of the GS/E's tidal debris. 

Another concern is that 5 out of 8 GCs have positive $L_z$. As discussed in \citet{wrinkles}, the GS/E debris cloud has an apparent tilt in the $E, L_z$ space such that the higher energy stars show net prograde motion. The net prograde motion of the eight suggested low-energy GCs is counter to this trend. While the high-energy GS/E GCs all have high orbital eccentricity, i.e. $0.8<e<1$, the additional low-energy candidate objects have significantly lower and more varied eccentricities, i.e. $0.3<e<0.8$. Unfortunately, we have chemical information only for 2 out of 8 clusters and these particular objects are both consistent with being a part of the in-situ population.

Finally, nothing makes the 14 GCs with uncertain progenitor (marked with orange diamonds) stand out from the rest of the in-situ clusters. These span a very broad range of $E$, $L_z$ and $L_{\perp}$. Chemical information is available for only one object from this group and it places it in the in-situ dominated region.

\citet{Malhan.etal.2022} presented classification of the Milky Way's GCs and streams using estimates of their total energy and actions using Gaia EDR3 kinematic measurements. All but one of the accreted structures these authors identify lie above the in-situ/accreted boundary we use and thus would also be classified as accreted by our method. One of their identified systems, Pontus, lies just below our classification boundary in the in-situ region. We note, however, that as shown by \citet{Dillamore.etal.2022} dynamical effects of the Milky Way bar can create horizontal clustering of stars and other dynamical traces in the general vicinity of the $E-L_z$ region where Pontus is identified. It remains to be seen whether chemical abundances of this system are consistent with its accreted or in-situ origin.  

\cite{Sun.etal.2023} presented a classification scheme for in-situ and accreted GCs that largely follows the approach of \citet{Massari2019}. In particular, similarly to \citet{Massari2019} these authors identify the in-situ GCs using ``disc'' and ``bulge'' populations but defined using a different set of criteria involving spatial and kinematic properties from the Gaia DR3 measurements by \cite{Baumgardt_Vasiliev2021}. These criteria identify $45.3\%$ GCs as formed in-situ and $38.4\%$ as accreted, with the remaining $16.3\%$ were deemed to have uncertain origin. Thus, although the approach is similar to \citet{Massari2019} different criteria used to identify in-situ clusters resulted in a higher in-situ GC fraction. The biggest difference between the \citet{Sun.etal.2023} and our classifications is in that the former assigns ``Kraken'' low-energy clusters to the accreted component. As we discussed above, however, there is no clear evidence that these clusters are a distinct grouping that can be clearly associated with an accretion event. 

\begin{figure}
  \centering
  \includegraphics[width=0.5\textwidth]{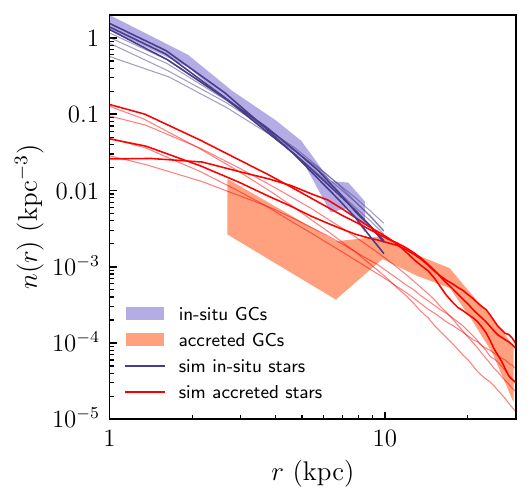}
  \caption[]{Number density of in-situ (blue) and accreted (red) GCs as a function of their galactocentric distance $r$. The shaded regions show the scatter around median profiles in the bootstrap samples of the MW GCs, while the blue and red lines show the number density of stellar particles from the FIRE-2 simulations of the MW-sized hosts (namely, objects {\tt m12f, m12i, m12m, m12w, m12r, m12b, m12c}). The in-situ and accreted stellar particles are weighted such that their metallicity distribution matches that of the in-situ and accreted MW GCs, as described in the text. The three objects, {\tt m12f, m12r, m12b}, that are closest in age-metallicity relation to the MW GCs \citep[see Fig. 13 in][]{Belokurov.Kravtsov.2023} are shown by the stronger lines, while the other objects are shown by the thinner lines.}
   \label{fig:nr_sim_comp}
\end{figure}
%
\subsection{Comparison with models and implications for GC formation}
\label{sec:gcform}

While detailed comparisons with models of GC formation are beyond the scope of this study, here we will discuss general comparisons focusing on the fraction of accreted clusters estimated in our classification and in the models. We will also present comparisons with statistics of the in-situ and accreted stellar particles in the FIRE-2 simulations of the MW-sized haloes \citep{Hopkins.etal.2018,Wetzel.etal.2023} and discuss the implications of these comparisons for models of GC formation and evolution. 

As we noted above, $\approx 1/3$ of surviving GCs in our classification are accreted. This is lower than in some of the recent models of GC formation. For example, the model of \citet{Chen_Gnedin2022} predicts for MW-sized hosts the ratio of the number of accreted to in-situ {\it surviving} GCs of $\sim 2/1$ to $3/1$. The number of in-situ GCs that form in the MW progenitor in their model is actually larger than the number of accreted clusters that ever formed, but many more in-situ clusters get tidally disrupted compared to the accreted clusters and the number of surviving clusters is thus dominated by the accreted GCs. 
The results of the model are thus quite sensitive to how tidal disruption of clusters is modelled. 
For example, in the previous version of this model \citep[see Fig. 6 in][]{Choksi.Gnedin.2019} with a different disruption model predicted population of the surviving GCs was dominated by the in-situ clusters. 

E-MOSAIC GC formation model predicts that the mean $52\pm 1\%$ of the surviving GCs were born in situ \citep{Keller.etal.2020}, although further analyses showed that the {\it ex-situ} fractions vary non-negligibly from object to object and have a range of $\approx 37\pm 11\%$ which is similar to the fraction in our classification \citep[][]{Kruijssen.etal.2019b,Trujillo_Gomez.etal.2021}. 
\citet{Trujillo_Gomez.etal.2023}, on the other hand, find median accreted fraction of $\approx 60\%$ in the same E-MOSAIC model, but with substantial scatter around it; the different fractions in different E-MOSAIC analyses are due to different selection of GC samples indicating sensitivity of the accreted fraction to details of selection. 
Generally, the accreted fraction is expected to have a significant scatter due to different assembly histories of objects of the same halo mass. 
Their model also predicts that the fraction of surviving clusters is approximately the same among accreted and in-situ clusters. We thus see a significant variation among GC models in what they predict for the accreted and in-situ GC populations and their survival.  

Overall, galaxy formation models predict that the accreted fraction of stellar population is very small in galaxies of $M_\star\lesssim 10^{10}\, M_\odot$ but increases rapidly for larger masses reaching accreted fractions of $\approx 20-50\%$ for galaxies with $M_\star\approx 10^{11}\, M_\odot$ \citep[e.g.,][]{Qu.etal.2017, Pilepich.etal.2018,Clauwens.etal.2018,Davison.etal.2020}. However, these fractions 
refer to the total masses of accreted and in-situ populations at all radii. 

As we saw, our classification implies that GC population within galactocentric distance of 10 kpc is dominated by in-situ clusters. We estimated the mass fraction of accreted stellar particles within such galactocentric distance in the seven MW-sized objects from the FIRE-2 simulation suite \citep{Hopkins.etal.2018,Wetzel.etal.2023} and find that $f_{\rm acc}(<10\,{\rm kpc})=m_{\star,\rm acc}/{m_{\star,\rm tot}}$ ranges from 2 to 7\% in 5 out of 7 galaxies, and reaches 13\% and 25\% in the other two systems. 
Overall, therefore, simulations predict that stellar population of MW-sized galaxies within the central 10 kpc is dominated by the in-situ stars. 

We have carried out another comparison, which is aimed to be more directly related to the accreted fraction of GCs. Namely, we examined distributions of several properties of stellar particles in the FIRE-2 simulations of the MW-sized galaxies {\tt m12b, m12c, m12f, m12i, m12m, m12r, and m12w} but weighted or selected so as to match metallicity distributions of the in-situ and accreted MW GCs. Figure~\ref{fig:nr_sim_comp} shows a comparison of the radial number density profiles of in-situ and accreted stellar particles in the FIRE-2 objects weighted in this way to the number density profiles of the in-situ and accreted GCs in our classification. The density profiles of stellar particles constructed this way are normalized so that the number density profile of the in-situ particles approximately matches the number density profile of in-situ in amplitude. The same normalization factor is used for both the in-situ and accreted stellar particles. 

\begin{figure*}
  \centering
  \includegraphics[width=\textwidth]{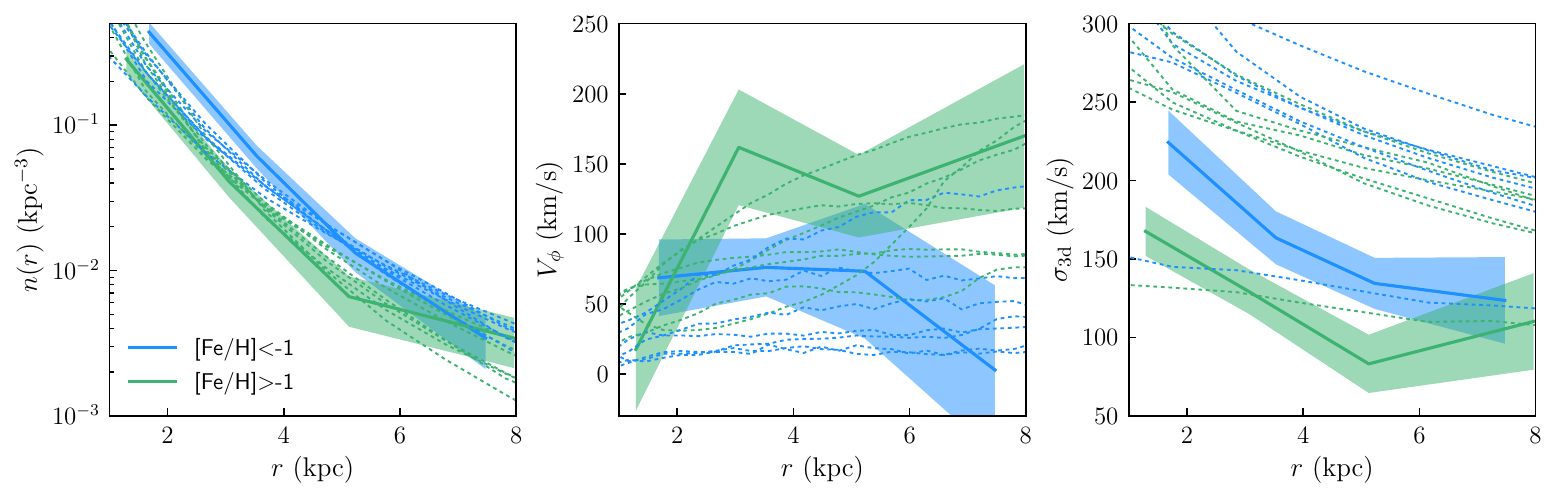}
  \caption[]{The radial profiles of the in-situ MW GCs of low ($\rm [Fe/H]<-1$) and high ($\rm [Fe/H]>-1$) metallicity shown in Figure~\ref{fig:hilo_pro} above compared with the corresponding profiles of in-situ stellar particles from the FIRE-2 simulations of the MW-sized galaxies of the same metallicity ranges. In computing these in-situ stellar particles are weighted such that their metallicity distribution matches that of the in-situ MW GCs, as described in the text. The figure shows that the in-situ stellar particles in simulations match the radial density profile and profile of the tangential velocity, as well as their changes between low- and high-metallicity samples. The trend is also reproduced for the velocity dispersion, but velocity dispersions in simulations are generally considerably higher than for MW GCs, except for the dispersion in the object {\tt m12r}. 
  }
   \label{fig:hilo_pro_sim}
\end{figure*}

Figure~\ref{fig:nr_sim_comp} shows that the number density profiles of in-situ and accreted stellar particles match the relative amplitude and shapes of the corresponding density profiles of the MW  GCs quite well. The match is especially good for three objects -- {\tt m12f, m12r, m12b} -- that have in-situ age-$\rm [Fe/H]$ stellar sequences closest to the corresponding sequence of in-situ GCs \citep[see Fig. 13 in][]{Belokurov.Kravtsov.2023}. The in-situ star particles are more centrally concentrated, while accreted particles have a much more extended distribution.
This is generally found in models of GC evolution \citep{Reina.Campos.etal2022b,Chen_Gnedin2022}, but here we see a remarkable match of both shapes and relative amplitudes of the observed profiles.

Figure~\ref{fig:hilo_pro_sim} shows comparisons of the number density, median tangential velocity $V_\phi$, and 3d velocity dispersion $\sigma_{\rm 3d}$ profiles of in-situ stellar particles with $\rm [Fe/H]<-1$ and $\rm [Fe/H]>-1$ with the corresponding profiles of the in-situ MW GCs. The metallicity trend in the number density profile is reproduced quite well. The metallicity trend in the $V_\phi$ profile is also qualitatively reproduced. 

The match of the relative amplitude and shape of the GC number density profiles of in-situ and accreted GCs by the stellar particle number density profiles in Figure~\ref{fig:nr_sim_comp} indicates that simulations capture realistically formation of stars and their dynamics. Assuming this is the case, the match {\it implies that GC formation is a part of regular star formation in the MW progenitor}. However, given that the metallicity distribution of GCs is different from that of the MW in-situ stars, GC formation was confined only to certain periods of the Galaxy evolution. These periods likely reflected periods of high gas accretion either when MW progenitor halo was still in the fast accretion regime or during spikes in the gas accretion rate in the slow accretion regime. One of such spikes could have been associated with the GS/E merger $\approx 10$ Gyr ago. This merger proceeded for a while and could have affected formation of stars and GCs with metallicities between $\rm [Fe/H]\approx -1.5$ and $-0.5$.

The good match of the observed GC and simulated stellar number density profiles also implies that disruption of GCs should not have a strong distance dependence, otherwise radial distribution of GCs would be different than the radial stellar particles that are not subject to disruption. This is in general agreement with models of GC formation and evolution \citep{Keller.etal.2020,Gieles2023}. Likewise, the agreement of the in-situ number density profiles at different metallicities indicates that tidal disruption should not have a strong metallicity dependence, which agrees with the model results of \citep[see Fig. 15][and O. Gnedin, priv.communication]{Keller.etal.2020}.

Recently, there has been a number of efforts to include explicitly the formation and evolution of massive gas clumps in high-resolution hydro-dynamical  simulations of the Milky Way disc. For example, \citet{Clarke2019} show that $\sim100$ of well-resolved gas clumps with masses between $3\times10^7 M_{\odot}$ and $10^{10} M_{\odot}$ can form in the early Milky Way. Most massive of these can sustain prolonged star formation and migrate through the inner regions of the Galaxy leaving a distinct imprint on the disc's chemical, structural and kinematic behaviour \citep[see also][]{Beraldo2020,Debattista2023,Garver2023}. The massive, early formed gas clumps described in the above models appear to be a natural progenitor of the population of Galactic disc's globular clusters discussed here.

It is interesting to note that mass of the stellar halo of our Galaxy is only $1.4\pm 0.4\times 10^9\, M_\odot$ \citep{Deason.etal.2019}, while in-situ stellar mass of our galaxy is $\approx 6\times 10^{10}\, M_\odot$ \citep[e.g.,][]{Licquia.Newman.2015}.  The overall fraction of accreted stars in our Galaxy is thus $f_{\rm acc}\approx 2\%$, while the fraction of accreted GCs in our classification by number is much larger: $58/107\approx 54\%$. By mass, the mass fraction in surviving accreted GCs is $\approx 40\%$, although if we estimate the mass fraction of accreted clusters using the initial GCs masses in the in-situ and accreted clusters estimated by \citet{Baumgardt2003} the initial accreted mass fraction is $\approx 20\%$. Regardless of how we estimate the accreted fraction, it is at least ten times larger than the overall accreted mass fraction in the MW's stellar population. 

It is likely that this is due to a combination of factors. First, MW stopped forming in-situ GCs $\gtrsim 10$ Gyrs ago (see ages in Fig. 3 above), while it continued to accrete GCs and form in-situ stars. Second, the number of GCs scales almost linearly with halo mass over more than five orders of magnitude in galaxy stellar mass with $\eta_{\rm GC}=M_{\rm GCs}/M_{\rm h}\approx 3\times 10^{-5}$ \citep[see][]{Spitler.Forbes.2009,Hudson.etal.2014,Harris.etal.2017,Forbes.etal.2018,Dornan.Harris.2023}, although it is somewhat uncertain at the smallest masses and may deviate from linearity in that regime \citep{Bastian2020,DeLucia.etal.2023}. Galaxy stellar mass scales, on the other hand, scales non-linearly in the dwarf galaxy regime \citep[e.g.,][]{Nadler.etal.2020}. Thus, dwarf galaxies bring proportionally more GCs compared to stars when they merge with the MW.

\subsection{Caveats}
\label{sec:caveats}

As we noted before, the use of a categorical boundary almost certainly will misclassify some accreted objects with energies below the boundary and vice versa. This needs to be kept in mind when considering individual GCs. Eventually, as reliable chemical element ratios and orbital parameter estimates become available for more clusters, it should be possible to refine the classification presented here. For example, the $\omega$ Centauri and NGC clusters are classified as in-situ clusters by our method but are likely to be remnant nuclear star clusters of accreted galaxies due to their large metallicity spread \citep[e.g.,][]{Pfeffer.etal.2021}. We mark the clusters that have some indications of being misclassified in comments in Table~\ref{tab:class} in Appendix~\ref{app:class_tab}.

Overall, our estimates indicate that the fraction of accreted clusters among those we classified as in-situ is likely $\lesssim 10\%$ (or fewer than 15 clusters). First, we do not see evidence of a significant sub-population of these clusters with distinct chemical and orbital properties. Second, the number density profiles of in-situ and accreted stellar particles that match the corresponding profiles of the GCs in our classification have relative amplitudes that correspond to $\lesssim 10\%$ of accreted stellar particles at $r<10$ kpc. 
Third, detailed analyses of galaxy formation simulations indicate that MW's halo and disc form earlier than most objects of similar mass \citep[][]{Mccluskey2023,Semenov.etal.2023a,Dillamore.etal.2023b}. \citet{Dillamore.etal.2023b} recently showed that such systems also have smaller than average fraction of accreted stars. Thus, it is very unlikely that the fraction of accreted GCs in the MW is high, especially in the central 10 kpc from the Galaxy centre.

\section{Summary and conclusions}
\label{sec:conc}

We use the $\rm [Al/Fe]$-calibrated in-situ/accreted classification in the $E-L_z$ plane introduced in \citet{Belokurov.Kravtsov.2023} to demonstrate that such classification results in the GC populations with distinct spatial, kinematic and chemical abundance distributions. The specific results presented in this paper and their implications are as follows.

\begin{itemize}
     \item[(i)] Figure~\ref{fig:spatial} shows that our classification results in GC samples with qualitatively different spatial distributions. In-situ clusters are located mainly at $\lesssim 10$ kpc from the centre of the Galaxy, while accreted clusters are mainly located at larger distances. The distribution of accreted clusters is almost spherical, while in-situ clusters are distributed in a flattened configuration aligned with the MW disc. \\[-2mm]

     \item[(ii)] Our classification splits the clusters into two distinct sequences in the age-metallicity plane (Fig.~\ref{fig:agefeh}) with in-situ GCs tracing the evolution of metallicity as a function of time of our Galaxy.\\[-2mm] 

     \item[(iii)] The accreted and in-situ clusters have different distributions of metallicities (Fig.~\ref{fig:feh_dist}). Most accreted clusters have $\rm [Fe/H]\lesssim -1$ and distribution of metallicities has a single peak at $[\rm Fe/H]\approx -1.6$. Metallicity distribution of the in-situ clusters spans a much wider range of $[-2.3,0]$ and has two peaks centered at $[\rm Fe/H]\approx -1.4$ and 
     $[\rm Fe/H]\approx -0.7$. The weak bi-modality of the overall metallicity distribution of the MW GCs is thus entirely due to the in-situ clusters.\\[-2mm] 
     
    \item[(iv)] We show that the in-situ GCs in our classification show a clear disc spin-up signature -- the increase of median $V_\phi$ at metallicities $\rm [Fe/H]\approx -1.3\div -1$ similar to the signature exhibited by the in-situ stars of the Milky Way.\\[-2mm] 
    
    \item[(v)] This feature signals MW's disc formation and the fact that it is also present in the kinematics of the in-situ GCs means that GCs with metallicities of $\rm [Fe/H]\gtrsim -1.3$ were born in the Milky Way disc, while lower metallicity GCs were born during early, turbulent, pre-disc stages of the evolution of the Galaxy and are part of the Aurora stellar component of the Milky Way. \\[-2mm]

    \item[(vi)] Ages and metallicities of in-situ GCs and the spin-up metallicity range indicate that MW's disc formed $\approx 11.7-12.7$ Gyrs ago or at $z\approx 3.1-5.3$.
    
    \item[(vii)] We explicitly show radial and velocity distributions of the Aurora clusters and higher metallicity in-situ clusters are different (Section~\ref{sec:lohicomp} and Fig.~\ref{fig:hilo_pro}).\\[-2mm]  

    \item[(viii)] We show that the accreted and in-situ GCs are well separated in the plane of $\rm [Al/Fe]-[Mg/Fe]$ abundance ratios.\\[-2mm]

     \item[(ix)] We show that the radial distribution of the in-situ and accreted GCs is very similar to the radial distribution of the in-situ and accreted stellar particles in the FIRE-2 galaxy formation simulations if particles are selected to have metallicity distribution similar to that of the MW GCs. This indicates that MW globular clusters are born as part of the normal star formation in the MW progenitor but during epochs most conducive for their formation.\\[-4mm] 
\end{itemize}

The classification method presented in this paper is meant to be applicable broadly to the entire GC population of the Milky Way. It is based only on the total energy and $L_z$ angular momentum because these are some of the very few quantities that are available for the entire GC sample. It is clear thus that the method is unlikely to be 100\% accurate. 
Nevertheless, we estimate that not more than $\approx 10\%$ of the clusters classified as in-situ in our method may actually be accreted. This classification can of course be refined further using additional formation for individual clusters, such as $\rm [Al/Fe]$ and $\rm [Mg/Fe]$ abundance ratios, as it becomes available. For example, $\omega$ Centauri and NGC 6273 clusters are likely misclassified by our method as in-situ, given the evidence for large metallicity spread in these systems which implies that they have likely been nuclear star clusters in accreted galaxies \citep[e.g.,][]{Pfeffer.etal.2021}. We indicate GCs that may be misclassified by our method in the comments column of Table~\ref{tab:class}, in which our classification for individual clusters is presented. 

The presented classification should be useful for testing models of globular cluster formation in the cosmological context. We stress, however, that recent analyses of galaxy formation simulations in comparisons with the kinematics of the in-situ stars of the Milky Way indicate that MW's halo and disc form earlier than most objects of similar mass \citep[][]{Aurora, Mccluskey2023,Semenov.etal.2023a,Dillamore.etal.2023b}. \citet{Dillamore.etal.2023b} recently showed that galaxies that undergo a GS/E-like 
merger and which form disc as early as the Milky Way have much smaller than average fractions of accreted stars. Thus, care should be taken when comparing models with specific MW GC sample and its accreted and in-situ subpopulations.

\section*{Acknowledgments}

We are grateful to Holger Baumgardt and Oleg Gnedin for useful discussions and to Eugene Vasiliev, Davide Massari, Marta Reina-Campos, Stephanie Monty for their comments that helped improve the quality of this manuscript.
AK was supported by the National Science Foundation grants AST-1714658 and AST-1911111 and NASA ATP grant 80NSSC20K0512.
This research made use of data from the European Space Agency mission Gaia
(\url{http://www.cosmos.esa.int/gaia}), processed by the Gaia Data
Processing and Analysis Consortium (DPAC,
\url{http://www.cosmos.esa.int/web/gaia/dpac/consortium}). Funding for the
DPAC has been provided by national institutions, in particular the
institutions participating in the Gaia Multilateral Agreement. 

This
paper made used of the Whole Sky Database (wsdb) created by Sergey
Koposov and maintained at the Institute of Astronomy, Cambridge with
financial support from the Science \& Technology Facilities Council
(STFC) and the European Research Council (ERC). We also used FIRE-2 simulation public data  \citep[][\url{http://flathub.flatironinstitute.org/fire}]{Wetzel.etal.2023}, which are part of the Feedback In Realistic Environments (FIRE) project, generated using the Gizmo code \citep{hopkins15} and the FIRE-2 physics model \citep{hopkins_etal18}. 
Analyses presented in this paper were greatly aided by the following free software packages: {\tt NumPy} \citep{NumPy}, {\tt SciPy} \citep{scipy}, {\tt Matplotlib} \citep{matplotlib}, and {\tt Scikit-learn} \citep{sklearn}. We have also used the Astrophysics Data Service (\href{http://adsabs.harvard.edu/abstract_service.html}{\tt ADS}) and \href{https://arxiv.org}{\tt arXiv} preprint repository extensively during this project and the writing of the paper.

\section*{Data Availability}

This study uses  \verb|allStarLite-dr17-synspec_rev1| and \verb|apogee_astroNN-DR17| catalogues publicly available at \url{https://www.sdss.org/dr17/irspec/spectro_data/}. The catalog of the MW globular clusters with distances used in this study is publicly available at
\url{https://people.smp.uq.edu.au/HolgerBaumgardt/globular/}. 
The FIRE-2 simulations used in this study are available at 
\url{http://flathub.flatironinstitute.org/fire}.

\bibliography{references}

\begin{thebibliography}{}
\makeatletter
\relax
\def\mn@urlcharsother{\let\do\@makeother \do\$\do\&\do\#\do\^\do\_\do\%\do\~}
\def\mn@doi{\begingroup\mn@urlcharsother \@ifnextchar [ {\mn@doi@}
  {\mn@doi@[]}}
\def\mn@doi@[#1]#2{\def\@tempa{#1}\ifx\@tempa\@empty \href
  {http://dx.doi.org/#2} {doi:#2}\else \href {http://dx.doi.org/#2} {#1}\fi
  \endgroup}
\def\mn@eprint#1#2{\mn@eprint@#1:#2::\@nil}
\def\mn@eprint@arXiv#1{\href {http://arxiv.org/abs/#1} {{\tt arXiv:#1}}}
\def\mn@eprint@dblp#1{\href {http://dblp.uni-trier.de/rec/bibtex/#1.xml}
  {dblp:#1}}
\def\mn@eprint@#1:#2:#3:#4\@nil{\def\@tempa {#1}\def\@tempb {#2}\def\@tempc
  {#3}\ifx \@tempc \@empty \let \@tempc \@tempb \let \@tempb \@tempa \fi \ifx
  \@tempb \@empty \def\@tempb {arXiv}\fi \@ifundefined
  {mn@eprint@\@tempb}{\@tempb:\@tempc}{\expandafter \expandafter \csname
  mn@eprint@\@tempb\endcsname \expandafter{\@tempc}}}

\bibitem[\protect\citeauthoryear{{Adamo} et~al.,}{{Adamo}
  et~al.}{2020}]{Adamo.etal.2020rev}
{Adamo} A.,  et~al., 2020, \mn@doi [\ssr] {10.1007/s11214-020-00690-x}, \href
  {https://ui.adsabs.harvard.edu/abs/2020SSRv..216...69A} {216, 69}

\bibitem[\protect\citeauthoryear{Ankerst, Breunig, Kriegel  \& Sander}{Ankerst
  et~al.}{1999}]{OPTICS}
Ankerst M.,  Breunig M.~M.,  Kriegel H.-P.,   Sander J.,  1999, \mn@doi [ACM
  SIGMOD Record] {http://doi.acm.org/10.1145/304181.304187}, 28, 49

\bibitem[\protect\citeauthoryear{{Armandroff}}{{Armandroff}}{1989}]{Armandroff.1989}
{Armandroff} T.~E.,  1989, \mn@doi [\aj] {10.1086/114988}, \href
  {https://ui.adsabs.harvard.edu/abs/1989AJ.....97..375A} {97, 375}

\bibitem[\protect\citeauthoryear{{Ashman} \& {Zepf}}{{Ashman} \&
  {Zepf}}{1992}]{Ashman.Zepf.1992}
{Ashman} K.~M.,  {Zepf} S.~E.,  1992, \mn@doi [\apj] {10.1086/170850}, \href
  {https://ui.adsabs.harvard.edu/abs/1992ApJ...384...50A} {384, 50}

\bibitem[\protect\citeauthoryear{{Ashman} \& {Zepf}}{{Ashman} \&
  {Zepf}}{2001}]{Ashman.Zepf.2001}
{Ashman} K.~M.,  {Zepf} S.~E.,  2001, \mn@doi [\aj] {10.1086/323133}, \href
  {https://ui.adsabs.harvard.edu/abs/2001AJ....122.1888A} {122, 1888}

\bibitem[\protect\citeauthoryear{{Barbuy} et~al.,}{{Barbuy} et~al.}{2016}]{hp1}
{Barbuy} B.,  et~al., 2016, \mn@doi [\aap] {10.1051/0004-6361/201628106}, \href
  {https://ui.adsabs.harvard.edu/abs/2016A&A...591A..53B} {591, A53}

\bibitem[\protect\citeauthoryear{{Barbuy}, {Chiappini}  \& {Gerhard}}{{Barbuy}
  et~al.}{2018a}]{Barbuy.etal.2018}
{Barbuy} B.,  {Chiappini} C.,   {Gerhard} O.,  2018a, \mn@doi [\araa]
  {10.1146/annurev-astro-081817-051826}, \href
  {https://ui.adsabs.harvard.edu/abs/2018ARA&A..56..223B} {56, 223}

\bibitem[\protect\citeauthoryear{{Barbuy} et~al.,}{{Barbuy}
  et~al.}{2018b}]{ngc6558}
{Barbuy} B.,  et~al., 2018b, \mn@doi [\aap] {10.1051/0004-6361/201833953},
  \href {https://ui.adsabs.harvard.edu/abs/2018A&A...619A.178B} {619, A178}

\bibitem[\protect\citeauthoryear{{Bastian}, {Pfeffer}, {Kruijssen}, {Crain},
  {Trujillo-Gomez}  \& {Reina-Campos}}{{Bastian} et~al.}{2020}]{Bastian2020}
{Bastian} N.,  {Pfeffer} J.,  {Kruijssen} J.~M.~D.,  {Crain} R.~A.,
  {Trujillo-Gomez} S.,   {Reina-Campos} M.,  2020, \mn@doi [\mnras]
  {10.1093/mnras/staa2453}, \href
  {https://ui.adsabs.harvard.edu/abs/2020MNRAS.498.1050B} {498, 1050}

\bibitem[\protect\citeauthoryear{{Baumgardt}}{{Baumgardt}}{2017}]{Baumgardt2017}
{Baumgardt} H.,  2017, \mn@doi [\mnras] {10.1093/mnras/stw2488}, \href
  {https://ui.adsabs.harvard.edu/abs/2017MNRAS.464.2174B} {464, 2174}

\bibitem[\protect\citeauthoryear{{Baumgardt} \& {Hilker}}{{Baumgardt} \&
  {Hilker}}{2018}]{Baumgardt2018}
{Baumgardt} H.,  {Hilker} M.,  2018, \mn@doi [\mnras] {10.1093/mnras/sty1057},
  \href {https://ui.adsabs.harvard.edu/abs/2018MNRAS.478.1520B} {478, 1520}

\bibitem[\protect\citeauthoryear{{Baumgardt} \& {Makino}}{{Baumgardt} \&
  {Makino}}{2003}]{Baumgardt2003}
{Baumgardt} H.,  {Makino} J.,  2003, \mn@doi [\mnras]
  {10.1046/j.1365-8711.2003.06286.x}, \href
  {https://ui.adsabs.harvard.edu/abs/2003MNRAS.340..227B} {340, 227}

\bibitem[\protect\citeauthoryear{{Baumgardt} \& {Vasiliev}}{{Baumgardt} \&
  {Vasiliev}}{2021}]{Baumgardt_Vasiliev2021}
{Baumgardt} H.,  {Vasiliev} E.,  2021, \mn@doi [\mnras]
  {10.1093/mnras/stab1474}, \href
  {https://ui.adsabs.harvard.edu/abs/2021MNRAS.505.5957B} {505, 5957}

\bibitem[\protect\citeauthoryear{{Baumgardt}, {Sollima}  \&
  {Hilker}}{{Baumgardt} et~al.}{2020}]{Baumgardt2020}
{Baumgardt} H.,  {Sollima} A.,   {Hilker} M.,  2020, \mn@doi [\pasa]
  {10.1017/pasa.2020.38}, \href
  {https://ui.adsabs.harvard.edu/abs/2020PASA...37...46B} {37, e046}

\bibitem[\protect\citeauthoryear{{Baumgardt}, {H{\'e}nault-Brunet}, {Dickson}
  \& {Sollima}}{{Baumgardt} et~al.}{2023}]{Baumgardt2023}
{Baumgardt} H.,  {H{\'e}nault-Brunet} V.,  {Dickson} N.,   {Sollima} A.,  2023,
  \mn@doi [\mnras] {10.1093/mnras/stad631}, \href
  {https://ui.adsabs.harvard.edu/abs/2023MNRAS.521.3991B} {521, 3991}

\bibitem[\protect\citeauthoryear{{Beasley}, {Baugh}, {Forbes}, {Sharples}  \&
  {Frenk}}{{Beasley} et~al.}{2002}]{Beasley.etal.2002}
{Beasley} M.~A.,  {Baugh} C.~M.,  {Forbes} D.~A.,  {Sharples} R.~M.,   {Frenk}
  C.~S.,  2002, \mn@doi [\mnras] {10.1046/j.1365-8711.2002.05402.x}, \href
  {https://ui.adsabs.harvard.edu/abs/2002MNRAS.333..383B} {333, 383}

\bibitem[\protect\citeauthoryear{{Belokurov} \& {Kravtsov}}{{Belokurov} \&
  {Kravtsov}}{2022}]{Aurora}
{Belokurov} V.,  {Kravtsov} A.,  2022, \mn@doi [\mnras]
  {10.1093/mnras/stac1267}, \href
  {https://ui.adsabs.harvard.edu/abs/2022MNRAS.514..689B} {514, 689}

\bibitem[\protect\citeauthoryear{{Belokurov} \& {Kravtsov}}{{Belokurov} \&
  {Kravtsov}}{2023}]{Belokurov.Kravtsov.2023}
{Belokurov} V.,  {Kravtsov} A.,  2023, \mn@doi [\mnras, in press]
  {10.48550/arXiv.2306.00060}, \href
  {https://ui.adsabs.harvard.edu/abs/2023arXiv230600060B} {p. arXiv:2306.00060}

\bibitem[\protect\citeauthoryear{{Belokurov}, {Vasiliev}, {Deason}, {Koposov},
  {Fattahi}, {Dillamore}, {Davies}  \& {Grand}}{{Belokurov}
  et~al.}{2023}]{wrinkles}
{Belokurov} V.,  {Vasiliev} E.,  {Deason} A.~J.,  {Koposov} S.~E.,  {Fattahi}
  A.,  {Dillamore} A.~M.,  {Davies} E.~Y.,   {Grand} R. J.~J.,  2023, \mn@doi
  [\mnras] {10.1093/mnras/stac3436}, \href
  {https://ui.adsabs.harvard.edu/abs/2023MNRAS.518.6200B} {518, 6200}

\bibitem[\protect\citeauthoryear{{Beraldo e Silva}, {Debattista},
  {Khachaturyants}  \& {Nidever}}{{Beraldo e Silva} et~al.}{2020}]{Beraldo2020}
{Beraldo e Silva} L.,  {Debattista} V.~P.,  {Khachaturyants} T.,   {Nidever}
  D.,  2020, \mn@doi [\mnras] {10.1093/mnras/staa065}, \href
  {https://ui.adsabs.harvard.edu/abs/2020MNRAS.492.4716B} {492, 4716}

\bibitem[\protect\citeauthoryear{{Bland-Hawthorn} \&
  {Gerhard}}{{Bland-Hawthorn} \& {Gerhard}}{2016}]{Bland_Hawthorn.Gerhard.2016}
{Bland-Hawthorn} J.,  {Gerhard} O.,  2016, \mn@doi [\araa]
  {10.1146/annurev-astro-081915-023441}, \href
  {https://ui.adsabs.harvard.edu/abs/2016ARA&A..54..529B} {54, 529}

\bibitem[\protect\citeauthoryear{{Bragaglia}, {Carretta}, {Sollima}, {Donati},
  {D'Orazi}, {Gratton}, {Lucatello}  \& {Sneden}}{{Bragaglia}
  et~al.}{2015}]{ngc6139}
{Bragaglia} A.,  {Carretta} E.,  {Sollima} A.,  {Donati} P.,  {D'Orazi} V.,
  {Gratton} R.~G.,  {Lucatello} S.,   {Sneden} C.,  2015, \mn@doi [\aap]
  {10.1051/0004-6361/201526592}, \href
  {https://ui.adsabs.harvard.edu/abs/2015A&A...583A..69B} {583, A69}

\bibitem[\protect\citeauthoryear{{Brown}, {Wallerstein}  \& {Zucker}}{{Brown}
  et~al.}{1997}]{rup106}
{Brown} J.~A.,  {Wallerstein} G.,   {Zucker} D.,  1997, \mn@doi [\aj]
  {10.1086/118463}, \href
  {https://ui.adsabs.harvard.edu/abs/1997AJ....114..180B} {114, 180}

\bibitem[\protect\citeauthoryear{{Burkert}, {Brown}  \& {Truran}}{{Burkert}
  et~al.}{1996}]{Burkert.etal.1996}
{Burkert} A.,  {Brown} J.~H.,   {Truran} J.~W.,  1996, in {Burkert} A.,
  {Hartmann} D.~H.,   {Majewski} S.~A.,  eds,  Astronomical Society of the
  Pacific Conference Series Vol. 112, The History of the Milky Way and Its
  Satellite System. p.~121

\bibitem[\protect\citeauthoryear{{Carretta}}{{Carretta}}{2015}]{ngc2808}
{Carretta} E.,  2015, \mn@doi [\apj] {10.1088/0004-637X/810/2/148}, \href
  {https://ui.adsabs.harvard.edu/abs/2015ApJ...810..148C} {810, 148}

\bibitem[\protect\citeauthoryear{{Carretta} \& {Bragaglia}}{{Carretta} \&
  {Bragaglia}}{2023}]{ngc6388}
{Carretta} E.,  {Bragaglia} A.,  2023, \mn@doi [arXiv e-prints]
  {10.48550/arXiv.2307.05478}, \href
  {https://ui.adsabs.harvard.edu/abs/2023arXiv230705478C} {p. arXiv:2307.05478}

\bibitem[\protect\citeauthoryear{{Carretta} et~al.,}{{Carretta}
  et~al.}{2010}]{ngc6715}
{Carretta} E.,  et~al., 2010, \mn@doi [\aap] {10.1051/0004-6361/201014924},
  \href {https://ui.adsabs.harvard.edu/abs/2010A&A...520A..95C} {520, A95}

\bibitem[\protect\citeauthoryear{{Carretta}, {Lucatello}, {Gratton},
  {Bragaglia}  \& {D'Orazi}}{{Carretta} et~al.}{2011}]{ngc1851}
{Carretta} E.,  {Lucatello} S.,  {Gratton} R.~G.,  {Bragaglia} A.,   {D'Orazi}
  V.,  2011, \mn@doi [\aap] {10.1051/0004-6361/201117269}, \href
  {https://ui.adsabs.harvard.edu/abs/2011A&A...533A..69C} {533, A69}

\bibitem[\protect\citeauthoryear{{Carretta}, {Gratton}, {Bragaglia}, {D'Orazi}
  \& {Lucatello}}{{Carretta} et~al.}{2013a}]{ngc6121}
{Carretta} E.,  {Gratton} R.~G.,  {Bragaglia} A.,  {D'Orazi} V.,   {Lucatello}
  S.,  2013a, \mn@doi [\aap] {10.1051/0004-6361/201220470}, \href
  {https://ui.adsabs.harvard.edu/abs/2013A&A...550A..34C} {550, A34}

\bibitem[\protect\citeauthoryear{{Carretta} et~al.,}{{Carretta}
  et~al.}{2013b}]{ngc362}
{Carretta} E.,  et~al., 2013b, \mn@doi [\aap] {10.1051/0004-6361/201321905},
  \href {https://ui.adsabs.harvard.edu/abs/2013A&A...557A.138C} {557, A138}

\bibitem[\protect\citeauthoryear{{Carretta} et~al.,}{{Carretta}
  et~al.}{2014}]{ngc4833}
{Carretta} E.,  et~al., 2014, \mn@doi [\aap] {10.1051/0004-6361/201323321},
  \href {https://ui.adsabs.harvard.edu/abs/2014A&A...564A..60C} {564, A60}

\bibitem[\protect\citeauthoryear{{Chen} \& {Gnedin}}{{Chen} \&
  {Gnedin}}{2022}]{Chen_Gnedin2022}
{Chen} Y.,  {Gnedin} O.~Y.,  2022, \mn@doi [\mnras] {10.1093/mnras/stac1651},
  \href {https://ui.adsabs.harvard.edu/abs/2022MNRAS.514.4736C} {514, 4736}

\bibitem[\protect\citeauthoryear{{Chen} \& {Gnedin}}{{Chen} \&
  {Gnedin}}{2023}]{Chen_Gnedin2023}
{Chen} Y.,  {Gnedin} O.~Y.,  2023, \mn@doi [\mnras] {10.1093/mnras/stad1328},
  \href {https://ui.adsabs.harvard.edu/abs/2023MNRAS.522.5638C} {522, 5638}

\bibitem[\protect\citeauthoryear{{Choksi} \& {Gnedin}}{{Choksi} \&
  {Gnedin}}{2019}]{Choksi.Gnedin.2019}
{Choksi} N.,  {Gnedin} O.~Y.,  2019, \mn@doi [\mnras] {10.1093/mnras/stz2097},
  \href {https://ui.adsabs.harvard.edu/abs/2019MNRAS.488.5409C} {488, 5409}

\bibitem[\protect\citeauthoryear{{Choksi}, {Gnedin}  \& {Li}}{{Choksi}
  et~al.}{2018}]{Choksi2018}
{Choksi} N.,  {Gnedin} O.~Y.,   {Li} H.,  2018, \mn@doi [\mnras]
  {10.1093/mnras/sty1952}, \href
  {https://ui.adsabs.harvard.edu/abs/2018MNRAS.480.2343C} {480, 2343}

\bibitem[\protect\citeauthoryear{{Clarke} et~al.,}{{Clarke}
  et~al.}{2019}]{Clarke2019}
{Clarke} A.~J.,  et~al., 2019, \mn@doi [\mnras] {10.1093/mnras/stz104}, \href
  {https://ui.adsabs.harvard.edu/abs/2019MNRAS.484.3476C} {484, 3476}

\bibitem[\protect\citeauthoryear{{Clauwens}, {Schaye}, {Franx}  \&
  {Bower}}{{Clauwens} et~al.}{2018}]{Clauwens.etal.2018}
{Clauwens} B.,  {Schaye} J.,  {Franx} M.,   {Bower} R.~G.,  2018, \mn@doi
  [\mnras] {10.1093/mnras/sty1229}, \href
  {https://ui.adsabs.harvard.edu/abs/2018MNRAS.478.3994C} {478, 3994}

\bibitem[\protect\citeauthoryear{{Conroy} et~al.,}{{Conroy}
  et~al.}{2022}]{Conroy2022}
{Conroy} C.,  et~al., 2022, \mn@doi [arXiv e-prints]
  {10.48550/arXiv.2204.02989}, \href
  {https://ui.adsabs.harvard.edu/abs/2022arXiv220402989C} {p. arXiv:2204.02989}

\bibitem[\protect\citeauthoryear{{C{\^o}t{\'e}}, {Marzke}, {West}  \&
  {Minniti}}{{C{\^o}t{\'e}} et~al.}{2000}]{Cote.etal.2000}
{C{\^o}t{\'e}} P.,  {Marzke} R.~O.,  {West} M.~J.,   {Minniti} D.,  2000,
  \mn@doi [\apj] {10.1086/308709}, \href
  {https://ui.adsabs.harvard.edu/abs/2000ApJ...533..869C} {533, 869}

\bibitem[\protect\citeauthoryear{{C{\^o}t{\'e}}, {West}  \&
  {Marzke}}{{C{\^o}t{\'e}} et~al.}{2002}]{Cote.etal.2002}
{C{\^o}t{\'e}} P.,  {West} M.~J.,   {Marzke} R.~O.,  2002, \mn@doi [\apj]
  {10.1086/338670}, \href
  {https://ui.adsabs.harvard.edu/abs/2002ApJ...567..853C} {567, 853}

\bibitem[\protect\citeauthoryear{{Crestani}, {Alves-Brito}, {Bono}, {Puls}  \&
  {Alonso-Garc{\'\i}a}}{{Crestani} et~al.}{2019}]{ngc6723}
{Crestani} J.,  {Alves-Brito} A.,  {Bono} G.,  {Puls} A.~A.,
  {Alonso-Garc{\'\i}a} J.,  2019, \mn@doi [\mnras] {10.1093/mnras/stz1674},
  \href {https://ui.adsabs.harvard.edu/abs/2019MNRAS.487.5463C} {487, 5463}

\bibitem[\protect\citeauthoryear{{Das}, {Hawkins}  \& {Jofr{\'e}}}{{Das}
  et~al.}{2020}]{Das2020}
{Das} P.,  {Hawkins} K.,   {Jofr{\'e}} P.,  2020, \mn@doi [\mnras]
  {10.1093/mnras/stz3537}, \href
  {https://ui.adsabs.harvard.edu/abs/2020MNRAS.493.5195D} {493, 5195}

\bibitem[\protect\citeauthoryear{{Davison}, {Norris}, {Pfeffer}, {Davies}  \&
  {Crain}}{{Davison} et~al.}{2020}]{Davison.etal.2020}
{Davison} T.~A.,  {Norris} M.~A.,  {Pfeffer} J.~L.,  {Davies} J.~J.,   {Crain}
  R.~A.,  2020, \mn@doi [\mnras] {10.1093/mnras/staa1816}, \href
  {https://ui.adsabs.harvard.edu/abs/2020MNRAS.497...81D} {497, 81}

\bibitem[\protect\citeauthoryear{{De Lucia}, {Kruijssen}, {Trujillo-Gomez},
  {Hirschmann}  \& {Xie}}{{De Lucia} et~al.}{2023}]{DeLucia.etal.2023}
{De Lucia} G.,  {Kruijssen} J.~M.~D.,  {Trujillo-Gomez} S.,  {Hirschmann} M.,
  {Xie} L.,  2023, \mn@doi [arXiv e-prints] {10.48550/arXiv.2307.02530}, \href
  {https://ui.adsabs.harvard.edu/abs/2023arXiv230702530D} {p. arXiv:2307.02530}

\bibitem[\protect\citeauthoryear{{Deason}, {Belokurov}  \& {Sanders}}{{Deason}
  et~al.}{2019}]{Deason.etal.2019}
{Deason} A.~J.,  {Belokurov} V.,   {Sanders} J.~L.,  2019, \mn@doi [\mnras]
  {10.1093/mnras/stz2793}, \href
  {https://ui.adsabs.harvard.edu/abs/2019MNRAS.490.3426D} {490, 3426}

\bibitem[\protect\citeauthoryear{{Debattista} et~al.,}{{Debattista}
  et~al.}{2023}]{Debattista2023}
{Debattista} V.~P.,  et~al., 2023, \mn@doi [\apj] {10.3847/1538-4357/acbb00},
  \href {https://ui.adsabs.harvard.edu/abs/2023ApJ...946..118D} {946, 118}

\bibitem[\protect\citeauthoryear{{Dillamore}, {Belokurov}, {Font}  \&
  {McCarthy}}{{Dillamore} et~al.}{2022}]{Dillamore.etal.2022}
{Dillamore} A.~M.,  {Belokurov} V.,  {Font} A.~S.,   {McCarthy} I.~G.,  2022,
  \mn@doi [\mnras] {10.1093/mnras/stac1038}, \href
  {https://ui.adsabs.harvard.edu/abs/2022MNRAS.513.1867D} {513, 1867}

\bibitem[\protect\citeauthoryear{{Dillamore}, {Belokurov}, {Evans}  \&
  {Davies}}{{Dillamore} et~al.}{2023a}]{Dillamore.etal.2023}
{Dillamore} A.~M.,  {Belokurov} V.,  {Evans} N.~W.,   {Davies} E.~Y.,  2023a,
  \mn@doi [\mnras] {10.1093/mnras/stad2136}, \href
  {https://ui.adsabs.harvard.edu/abs/2023MNRAS.tmp.2042D} {}

\bibitem[\protect\citeauthoryear{{Dillamore}, {Belokurov}, {Kravtsov}  \&
  {Font}}{{Dillamore} et~al.}{2023b}]{Dillamore.etal.2023b}
{Dillamore} A.~M.,  {Belokurov} V.,  {Kravtsov} A.,   {Font} A.~S.,  2023b,
  arXiv e-prints, \href {https://ui.adsabs.harvard.edu/abs/2023arXiv230908658D}
  {p. arXiv:2309.08658}

\bibitem[\protect\citeauthoryear{{Dinescu}, {Girard}  \& {van
  Altena}}{{Dinescu} et~al.}{1999}]{Dinescu.etal.1999}
{Dinescu} D.~I.,  {Girard} T.~M.,   {van Altena} W.~F.,  1999, \mn@doi [\aj]
  {10.1086/300807}, \href
  {https://ui.adsabs.harvard.edu/abs/1999AJ....117.1792D} {117, 1792}

\bibitem[\protect\citeauthoryear{{Dornan} \& {Harris}}{{Dornan} \&
  {Harris}}{2023}]{Dornan.Harris.2023}
{Dornan} V.,  {Harris} W.~E.,  2023, \mn@doi [arXiv e-prints]
  {10.48550/arXiv.2304.11210}, \href
  {https://ui.adsabs.harvard.edu/abs/2023arXiv230411210D} {p. arXiv:2304.11210}

\bibitem[\protect\citeauthoryear{{Eggen}, {Lynden-Bell}  \& {Sandage}}{{Eggen}
  et~al.}{1962}]{Eggen.etal.1962}
{Eggen} O.~J.,  {Lynden-Bell} D.,   {Sandage} A.~R.,  1962, \mn@doi [\apj]
  {10.1086/147433}, \href
  {https://ui.adsabs.harvard.edu/abs/1962ApJ...136..748E} {136, 748}

\bibitem[\protect\citeauthoryear{{El-Badry}, {Quataert}, {Weisz}, {Choksi}  \&
  {Boylan-Kolchin}}{{El-Badry} et~al.}{2019}]{El_Badry_etal.2019}
{El-Badry} K.,  {Quataert} E.,  {Weisz} D.~R.,  {Choksi} N.,   {Boylan-Kolchin}
  M.,  2019, \mn@doi [\mnras] {10.1093/mnras/sty3007}, \href
  {https://ui.adsabs.harvard.edu/abs/2019MNRAS.482.4528E} {482, 4528}

\bibitem[\protect\citeauthoryear{{Fall} \& {Rees}}{{Fall} \&
  {Rees}}{1985}]{Fall.Rees.1985}
{Fall} S.~M.,  {Rees} M.~J.,  1985, \mn@doi [\apj] {10.1086/163585}, \href
  {https://ui.adsabs.harvard.edu/abs/1985ApJ...298...18F} {298, 18}

\bibitem[\protect\citeauthoryear{{Forbes}}{{Forbes}}{2020}]{Forbes2020}
{Forbes} D.~A.,  2020, \mn@doi [\mnras] {10.1093/mnras/staa245}, \href
  {https://ui.adsabs.harvard.edu/abs/2020MNRAS.493..847F} {493, 847}

\bibitem[\protect\citeauthoryear{{Forbes} \& {Bridges}}{{Forbes} \&
  {Bridges}}{2010}]{Forbes2010}
{Forbes} D.~A.,  {Bridges} T.,  2010, \mn@doi [\mnras]
  {10.1111/j.1365-2966.2010.16373.x}, \href
  {https://ui.adsabs.harvard.edu/abs/2010MNRAS.404.1203F} {404, 1203}

\bibitem[\protect\citeauthoryear{{Forbes}, {Read}, {Gieles}  \&
  {Collins}}{{Forbes} et~al.}{2018}]{Forbes.etal.2018}
{Forbes} D.~A.,  {Read} J.~I.,  {Gieles} M.,   {Collins} M. L.~M.,  2018,
  \mn@doi [\mnras] {10.1093/mnras/sty2584}, \href
  {https://ui.adsabs.harvard.edu/abs/2018MNRAS.481.5592F} {481, 5592}

\bibitem[\protect\citeauthoryear{{Gaia Collaboration} et~al.,}{{Gaia
  Collaboration} et~al.}{2016}]{Gaia}
{Gaia Collaboration} et~al., 2016, \mn@doi [\aap]
  {10.1051/0004-6361/201629272}, \href
  {https://ui.adsabs.harvard.edu/abs/2016A&A...595A...1G} {595, A1}

\bibitem[\protect\citeauthoryear{{Gaia Collaboration} et~al.,}{{Gaia
  Collaboration} et~al.}{2021}]{gaia_edr3}
{Gaia Collaboration} et~al., 2021, \mn@doi [\aap]
  {10.1051/0004-6361/202039657}, \href
  {https://ui.adsabs.harvard.edu/abs/2021A&A...649A...1G} {649, A1}

\bibitem[\protect\citeauthoryear{{Garver}, {Nidever}, {Debattista}, {Beraldo e
  Silva}  \& {Khachaturyants}}{{Garver} et~al.}{2023}]{Garver2023}
{Garver} B.~R.,  {Nidever} D.~L.,  {Debattista} V.~P.,  {Beraldo e Silva} L.,
  {Khachaturyants} T.,  2023, \mn@doi [\apj] {10.3847/1538-4357/acdfc6}, \href
  {https://ui.adsabs.harvard.edu/abs/2023ApJ...953..128G} {953, 128}

\bibitem[\protect\citeauthoryear{{Gieles} \& {Gnedin}}{{Gieles} \&
  {Gnedin}}{2023}]{Gieles2023}
{Gieles} M.,  {Gnedin} O.,  2023, \mn@doi [arXiv e-prints]
  {10.48550/arXiv.2303.03791}, \href
  {https://ui.adsabs.harvard.edu/abs/2023arXiv230303791G} {p. arXiv:2303.03791}

\bibitem[\protect\citeauthoryear{{Gunn}}{{Gunn}}{1980}]{Gunn.1980}
{Gunn} J.~E.,  1980, in {Hanes} D.,  {Madore} B.,  eds, Globular Clusters.
  p.~301

\bibitem[\protect\citeauthoryear{{Harris}}{{Harris}}{2010}]{Harris2010}
{Harris} W.~E.,  2010, \mn@doi [arXiv e-prints] {10.48550/arXiv.1012.3224},
  \href {https://ui.adsabs.harvard.edu/abs/2010arXiv1012.3224H} {p.
  arXiv:1012.3224}

\bibitem[\protect\citeauthoryear{{Harris} \& {Pudritz}}{{Harris} \&
  {Pudritz}}{1994}]{Harris.Pudritz.1994}
{Harris} W.~E.,  {Pudritz} R.~E.,  1994, \mn@doi [\apj] {10.1086/174310}, \href
  {https://ui.adsabs.harvard.edu/abs/1994ApJ...429..177H} {429, 177}

\bibitem[\protect\citeauthoryear{{Harris}, {Blakeslee}  \& {Harris}}{{Harris}
  et~al.}{2017}]{Harris.etal.2017}
{Harris} W.~E.,  {Blakeslee} J.~P.,   {Harris} G. L.~H.,  2017, \mn@doi [\apj]
  {10.3847/1538-4357/836/1/67}, \href
  {https://ui.adsabs.harvard.edu/abs/2017ApJ...836...67H} {836, 67}

\bibitem[\protect\citeauthoryear{{Hasselquist} et~al.,}{{Hasselquist}
  et~al.}{2021}]{Hasselquist2021}
{Hasselquist} S.,  et~al., 2021, \mn@doi [\apj] {10.3847/1538-4357/ac25f9},
  \href {https://ui.adsabs.harvard.edu/abs/2021ApJ...923..172H} {923, 172}

\bibitem[\protect\citeauthoryear{{Hawkins}, {Jofr{\'e}}, {Masseron}  \&
  {Gilmore}}{{Hawkins} et~al.}{2015}]{Hawkins2015}
{Hawkins} K.,  {Jofr{\'e}} P.,  {Masseron} T.,   {Gilmore} G.,  2015, \mn@doi
  [\mnras] {10.1093/mnras/stv1586}, \href
  {https://ui.adsabs.harvard.edu/abs/2015MNRAS.453..758H} {453, 758}

\bibitem[\protect\citeauthoryear{{Holtzman} et~al.,}{{Holtzman}
  et~al.}{1996}]{Holzman.etal.1996}
{Holtzman} J.~A.,  et~al., 1996, \mn@doi [\aj] {10.1086/118025}, \href
  {https://ui.adsabs.harvard.edu/abs/1996AJ....112..416H} {112, 416}

\bibitem[\protect\citeauthoryear{{Hopkins}}{{Hopkins}}{2015}]{hopkins15}
{Hopkins} P.~F.,  2015, \mn@doi [\mnras] {10.1093/mnras/stv195}, \href
  {https://ui.adsabs.harvard.edu/abs/2015MNRAS.450...53H} {450, 53}

\bibitem[\protect\citeauthoryear{{Hopkins} et~al.,}{{Hopkins}
  et~al.}{2018a}]{Hopkins.etal.2018}
{Hopkins} P.~F.,  et~al., 2018a, \mn@doi [\mnras] {10.1093/mnras/sty1690},
  \href {https://ui.adsabs.harvard.edu/abs/2018MNRAS.480..800H} {480, 800}

\bibitem[\protect\citeauthoryear{{Hopkins} et~al.,}{{Hopkins}
  et~al.}{2018b}]{hopkins_etal18}
{Hopkins} P.~F.,  et~al., 2018b, \mn@doi [\mnras] {10.1093/mnras/sty1690},
  \href {https://ui.adsabs.harvard.edu/abs/2018MNRAS.480..800H} {480, 800}

\bibitem[\protect\citeauthoryear{{Hudson}, {Harris}  \& {Harris}}{{Hudson}
  et~al.}{2014}]{Hudson.etal.2014}
{Hudson} M.~J.,  {Harris} G.~L.,   {Harris} W.~E.,  2014, \mn@doi [\apjl]
  {10.1088/2041-8205/787/1/L5}, \href
  {https://ui.adsabs.harvard.edu/abs/2014ApJ...787L...5H} {787, L5}

\bibitem[\protect\citeauthoryear{Hunter}{Hunter}{2007}]{matplotlib}
Hunter J.~D.,  2007, \mn@doi [Computing In Science \& Engineering]
  {10.1109/MCSE.2007.55}, 9, 90

\bibitem[\protect\citeauthoryear{{Johnson}, {Caldwell}, {Rich}  \&
  {Walker}}{{Johnson} et~al.}{2017a}]{ngc6229}
{Johnson} C.~I.,  {Caldwell} N.,  {Rich} R.~M.,   {Walker} M.~G.,  2017a,
  \mn@doi [\aj] {10.3847/1538-3881/aa86ac}, \href
  {https://ui.adsabs.harvard.edu/abs/2017AJ....154..155J} {154, 155}

\bibitem[\protect\citeauthoryear{{Johnson}, {Caldwell}, {Rich}, {Mateo},
  {Bailey}, {Clarkson}, {Olszewski}  \& {Walker}}{{Johnson}
  et~al.}{2017b}]{ngc6273}
{Johnson} C.~I.,  {Caldwell} N.,  {Rich} R.~M.,  {Mateo} M.,  {Bailey} John~I.
  I.,  {Clarkson} W.~I.,  {Olszewski} E.~W.,   {Walker} M.~G.,  2017b, \mn@doi
  [\apj] {10.3847/1538-4357/836/2/168}, \href
  {https://ui.adsabs.harvard.edu/abs/2017ApJ...836..168J} {836, 168}

\bibitem[\protect\citeauthoryear{{Johnson}, {Caldwell}, {Rich}, {Mateo},
  {Bailey}, {Olszewski}  \& {Walker}}{{Johnson} et~al.}{2017c}]{ngc5986}
{Johnson} C.~I.,  {Caldwell} N.,  {Rich} R.~M.,  {Mateo} M.,  {Bailey} John~I.
  I.,  {Olszewski} E.~W.,   {Walker} M.~G.,  2017c, \mn@doi [\apj]
  {10.3847/1538-4357/aa7414}, \href
  {https://ui.adsabs.harvard.edu/abs/2017ApJ...842...24J} {842, 24}

\bibitem[\protect\citeauthoryear{{Johnson}, {Rich}, {Caldwell}, {Mateo},
  {Bailey}, {Olszewski}  \& {Walker}}{{Johnson} et~al.}{2018}]{ngc6569}
{Johnson} C.~I.,  {Rich} R.~M.,  {Caldwell} N.,  {Mateo} M.,  {Bailey} John~I.
  I.,  {Olszewski} E.~W.,   {Walker} M.~G.,  2018, \mn@doi [\aj]
  {10.3847/1538-3881/aaa294}, \href
  {https://ui.adsabs.harvard.edu/abs/2018AJ....155...71J} {155, 71}

\bibitem[\protect\citeauthoryear{{Johnson}, {Caldwell}, {Michael Rich}, {Mateo}
   \& {Bailey}}{{Johnson} et~al.}{2019}]{ngc6402}
{Johnson} C.~I.,  {Caldwell} N.,  {Michael Rich} R.,  {Mateo} M.,   {Bailey}
  J.~I.,  2019, \mn@doi [\mnras] {10.1093/mnras/stz587}, \href
  {https://ui.adsabs.harvard.edu/abs/2019MNRAS.485.4311J} {485, 4311}

\bibitem[\protect\citeauthoryear{Jones, Oliphant, Peterson  et~al.}{Jones
  et~al.}{01  }]{scipy}
Jones E.,  Oliphant T.,  Peterson P.,   et~al., 2001--, {SciPy}: Open source
  scientific tools for {Python}, \url {http://www.scipy.org/}

\bibitem[\protect\citeauthoryear{{Kacharov}, {Koch}  \& {McWilliam}}{{Kacharov}
  et~al.}{2013}]{ngc6864}
{Kacharov} N.,  {Koch} A.,   {McWilliam} A.,  2013, \mn@doi [\aap]
  {10.1051/0004-6361/201321392}, \href
  {https://ui.adsabs.harvard.edu/abs/2013A&A...554A..81K} {554, A81}

\bibitem[\protect\citeauthoryear{{Keller}, {Kruijssen}, {Pfeffer},
  {Reina-Campos}, {Bastian}, {Trujillo-Gomez}, {Hughes}  \& {Crain}}{{Keller}
  et~al.}{2020}]{Keller.etal.2020}
{Keller} B.~W.,  {Kruijssen} J.~M.~D.,  {Pfeffer} J.,  {Reina-Campos} M.,
  {Bastian} N.,  {Trujillo-Gomez} S.,  {Hughes} M.~E.,   {Crain} R.~A.,  2020,
  \mn@doi [\mnras] {10.1093/mnras/staa1439}, \href
  {https://ui.adsabs.harvard.edu/abs/2020MNRAS.495.4248K} {495, 4248}

\bibitem[\protect\citeauthoryear{{Kravtsov} \& {Gnedin}}{{Kravtsov} \&
  {Gnedin}}{2005}]{Kravtsov2005}
{Kravtsov} A.~V.,  {Gnedin} O.~Y.,  2005, \mn@doi [\apj] {10.1086/428636},
  \href {https://ui.adsabs.harvard.edu/abs/2005ApJ...623..650K} {623, 650}

\bibitem[\protect\citeauthoryear{{Kravtsov}, {Vikhlinin}  \&
  {Meshcheryakov}}{{Kravtsov} et~al.}{2018}]{Kravtsov.etal.2018}
{Kravtsov} A.~V.,  {Vikhlinin} A.~A.,   {Meshcheryakov} A.~V.,  2018, \mn@doi
  [Astronomy Letters] {10.1134/S1063773717120015}, \href
  {https://ui.adsabs.harvard.edu/abs/2018AstL...44....8K} {44, 8}

\bibitem[\protect\citeauthoryear{{Kruijssen}}{{Kruijssen}}{2015}]{Kruijssen2015}
{Kruijssen} J.~M.~D.,  2015, \mn@doi [\mnras] {10.1093/mnras/stv2026}, \href
  {https://ui.adsabs.harvard.edu/abs/2015MNRAS.454.1658K} {454, 1658}

\bibitem[\protect\citeauthoryear{{Kruijssen}, {Pfeffer}, {Crain}  \&
  {Bastian}}{{Kruijssen} et~al.}{2019a}]{Kruijssen.etal.2019b}
{Kruijssen} J.~M.~D.,  {Pfeffer} J.~L.,  {Crain} R.~A.,   {Bastian} N.,  2019a,
  \mn@doi [\mnras] {10.1093/mnras/stz968}, \href
  {https://ui.adsabs.harvard.edu/abs/2019MNRAS.486.3134K} {486, 3134}

\bibitem[\protect\citeauthoryear{{Kruijssen}, {Pfeffer}, {Reina-Campos},
  {Crain}  \& {Bastian}}{{Kruijssen} et~al.}{2019b}]{Kruijssen.etal.2019}
{Kruijssen} J.~M.~D.,  {Pfeffer} J.~L.,  {Reina-Campos} M.,  {Crain} R.~A.,
  {Bastian} N.,  2019b, \mn@doi [\mnras] {10.1093/mnras/sty1609}, \href
  {https://ui.adsabs.harvard.edu/abs/2019MNRAS.486.3180K} {486, 3180}

\bibitem[\protect\citeauthoryear{{Kruijssen} et~al.,}{{Kruijssen}
  et~al.}{2020}]{Kruijssen2020}
{Kruijssen} J.~M.~D.,  et~al., 2020, \mn@doi [\mnras] {10.1093/mnras/staa2452},
  \href {https://ui.adsabs.harvard.edu/abs/2020MNRAS.498.2472K} {498, 2472}

\bibitem[\protect\citeauthoryear{{Krumholz}, {McKee}  \&
  {Bland-Hawthorn}}{{Krumholz} et~al.}{2019}]{Krumholz2019}
{Krumholz} M.~R.,  {McKee} C.~F.,   {Bland-Hawthorn} J.,  2019, \mn@doi [\araa]
  {10.1146/annurev-astro-091918-104430}, \href
  {https://ui.adsabs.harvard.edu/abs/2019ARA&A..57..227K} {57, 227}

\bibitem[\protect\citeauthoryear{{Lamb}, {Venn}, {Shetrone}, {Sakari}  \&
  {Pritzl}}{{Lamb} et~al.}{2015}]{ngc5466}
{Lamb} M.~P.,  {Venn} K.~A.,  {Shetrone} M.~D.,  {Sakari} C.~M.,   {Pritzl}
  B.~J.,  2015, \mn@doi [\mnras] {10.1093/mnras/stu2674}, \href
  {https://ui.adsabs.harvard.edu/abs/2015MNRAS.448...42L} {448, 42}

\bibitem[\protect\citeauthoryear{{Lapenna}, {Mucciarelli}, {Ferraro},
  {Origlia}, {Lanzoni}, {Massari}  \& {Dalessandro}}{{Lapenna}
  et~al.}{2015}]{ngc6266}
{Lapenna} E.,  {Mucciarelli} A.,  {Ferraro} F.~R.,  {Origlia} L.,  {Lanzoni}
  B.,  {Massari} D.,   {Dalessandro} E.,  2015, \mn@doi [\apj]
  {10.1088/0004-637X/813/2/97}, \href
  {https://ui.adsabs.harvard.edu/abs/2015ApJ...813...97L} {813, 97}

\bibitem[\protect\citeauthoryear{{Leaman}, {VandenBerg}  \& {Mendel}}{{Leaman}
  et~al.}{2013}]{Leaman2013}
{Leaman} R.,  {VandenBerg} D.~A.,   {Mendel} J.~T.,  2013, \mn@doi [\mnras]
  {10.1093/mnras/stt1540}, \href
  {https://ui.adsabs.harvard.edu/abs/2013MNRAS.436..122L} {436, 122}

\bibitem[\protect\citeauthoryear{{Lee}, {Carney}  \& {Habgood}}{{Lee}
  et~al.}{2005}]{ngc4590}
{Lee} J.-W.,  {Carney} B.~W.,   {Habgood} M.~J.,  2005, \mn@doi [\aj]
  {10.1086/426338}, \href
  {https://ui.adsabs.harvard.edu/abs/2005AJ....129..251L} {129, 251}

\bibitem[\protect\citeauthoryear{{Licquia} \& {Newman}}{{Licquia} \&
  {Newman}}{2015}]{Licquia.Newman.2015}
{Licquia} T.~C.,  {Newman} J.~A.,  2015, \mn@doi [\apj]
  {10.1088/0004-637X/806/1/96}, \href
  {https://ui.adsabs.harvard.edu/abs/2015ApJ...806...96L} {806, 96}

\bibitem[\protect\citeauthoryear{{Lindegren} et~al.,}{{Lindegren}
  et~al.}{2021}]{Lindegren2021}
{Lindegren} L.,  et~al., 2021, \mn@doi [\aap] {10.1051/0004-6361/202039709},
  \href {https://ui.adsabs.harvard.edu/abs/2021A&A...649A...2L} {649, A2}

\bibitem[\protect\citeauthoryear{{Malhan} et~al.,}{{Malhan}
  et~al.}{2022}]{Malhan.etal.2022}
{Malhan} K.,  et~al., 2022, \mn@doi [\apj] {10.3847/1538-4357/ac4d2a}, \href
  {https://ui.adsabs.harvard.edu/abs/2022ApJ...926..107M} {926, 107}

\bibitem[\protect\citeauthoryear{{Mar{\'\i}n-Franch}
  et~al.,}{{Mar{\'\i}n-Franch} et~al.}{2009}]{Marin_Franch.etal.2009}
{Mar{\'\i}n-Franch} A.,  et~al., 2009, \mn@doi [\apj]
  {10.1088/0004-637X/694/2/1498}, \href
  {https://ui.adsabs.harvard.edu/abs/2009ApJ...694.1498M} {694, 1498}

\bibitem[\protect\citeauthoryear{{Marino} et~al.,}{{Marino}
  et~al.}{2015}]{ngc5286}
{Marino} A.~F.,  et~al., 2015, \mn@doi [\mnras] {10.1093/mnras/stv420}, \href
  {https://ui.adsabs.harvard.edu/abs/2015MNRAS.450..815M} {450, 815}

\bibitem[\protect\citeauthoryear{{Marino} et~al.,}{{Marino}
  et~al.}{2019}]{ngc3201}
{Marino} A.~F.,  et~al., 2019, \mn@doi [\apj] {10.3847/1538-4357/ab53d9}, \href
  {https://ui.adsabs.harvard.edu/abs/2019ApJ...887...91M} {887, 91}

\bibitem[\protect\citeauthoryear{{Marino} et~al.,}{{Marino}
  et~al.}{2021}]{ngc1261}
{Marino} A.~F.,  et~al., 2021, \mn@doi [\apj] {10.3847/1538-4357/ac282c}, \href
  {https://ui.adsabs.harvard.edu/abs/2021ApJ...923...22M} {923, 22}

\bibitem[\protect\citeauthoryear{{Maschberger} \& {Kroupa}}{{Maschberger} \&
  {Kroupa}}{2007}]{Maschberger.Kroupa.2007}
{Maschberger} T.,  {Kroupa} P.,  2007, \mn@doi [\mnras]
  {10.1111/j.1365-2966.2007.11891.x}, \href
  {https://ui.adsabs.harvard.edu/abs/2007MNRAS.379...34M} {379, 34}

\bibitem[\protect\citeauthoryear{{Massari} et~al.,}{{Massari}
  et~al.}{2017}]{ngc6362}
{Massari} D.,  et~al., 2017, \mn@doi [\mnras] {10.1093/mnras/stx549}, \href
  {https://ui.adsabs.harvard.edu/abs/2017MNRAS.468.1249M} {468, 1249}

\bibitem[\protect\citeauthoryear{{Massari}, {Koppelman}  \& {Helmi}}{{Massari}
  et~al.}{2019}]{Massari2019}
{Massari} D.,  {Koppelman} H.~H.,   {Helmi} A.,  2019, \mn@doi [\aap]
  {10.1051/0004-6361/201936135}, \href
  {https://ui.adsabs.harvard.edu/abs/2019A&A...630L...4M} {630, L4}

\bibitem[\protect\citeauthoryear{{McCluskey}, {Wetzel}, {Loebman}, {Moreno}  \&
  {Faucher-Giguere}}{{McCluskey} et~al.}{2023}]{Mccluskey2023}
{McCluskey} F.,  {Wetzel} A.,  {Loebman} S.~R.,  {Moreno} J.,
  {Faucher-Giguere} C.-A.,  2023, \mn@doi [arXiv e-prints]
  {10.48550/arXiv.2303.14210}, \href
  {https://ui.adsabs.harvard.edu/abs/2023arXiv230314210M} {p. arXiv:2303.14210}

\bibitem[\protect\citeauthoryear{{Minelli}, {Mucciarelli}, {Massari},
  {Bellazzini}, {Romano}  \& {Ferraro}}{{Minelli} et~al.}{2021}]{Minelli2021}
{Minelli} A.,  {Mucciarelli} A.,  {Massari} D.,  {Bellazzini} M.,  {Romano} D.,
    {Ferraro} F.~R.,  2021, \mn@doi [\apjl] {10.3847/2041-8213/ac2156}, \href
  {https://ui.adsabs.harvard.edu/abs/2021ApJ...918L..32M} {918, L32}

\bibitem[\protect\citeauthoryear{{Montecinos}, {Villanova}, {Mu{\~n}oz}  \&
  {Cort{\'e}s}}{{Montecinos} et~al.}{2021}]{ngc6553}
{Montecinos} C.,  {Villanova} S.,  {Mu{\~n}oz} C.,   {Cort{\'e}s} C.~C.,  2021,
  \mn@doi [\mnras] {10.1093/mnras/stab712}, \href
  {https://ui.adsabs.harvard.edu/abs/2021MNRAS.503.4336M} {503, 4336}

\bibitem[\protect\citeauthoryear{{Monty} et~al.,}{{Monty}
  et~al.}{2023}]{Monty2023}
{Monty} S.,  et~al., 2023, \mn@doi [\mnras] {10.1093/mnras/stad1154}, \href
  {https://ui.adsabs.harvard.edu/abs/2023MNRAS.522.4404M} {522, 4404}

\bibitem[\protect\citeauthoryear{{Mu{\~n}oz} et~al.,}{{Mu{\~n}oz}
  et~al.}{2018}]{ngc6528}
{Mu{\~n}oz} C.,  et~al., 2018, \mn@doi [\aap] {10.1051/0004-6361/201833373},
  \href {https://ui.adsabs.harvard.edu/abs/2018A&A...620A..96M} {620, A96}

\bibitem[\protect\citeauthoryear{{Mura-Guzm{\'a}n}, {Villanova}, {Mu{\~n}oz}
  \& {Tang}}{{Mura-Guzm{\'a}n} et~al.}{2018}]{ngc5927}
{Mura-Guzm{\'a}n} A.,  {Villanova} S.,  {Mu{\~n}oz} C.,   {Tang} B.,  2018,
  \mn@doi [\mnras] {10.1093/mnras/stx2918}, \href
  {https://ui.adsabs.harvard.edu/abs/2018MNRAS.474.4541M} {474, 4541}

\bibitem[\protect\citeauthoryear{{Muratov} \& {Gnedin}}{{Muratov} \&
  {Gnedin}}{2010}]{Muratov.Gnedin.2010}
{Muratov} A.~L.,  {Gnedin} O.~Y.,  2010, \mn@doi [\apj]
  {10.1088/0004-637X/718/2/1266}, \href
  {https://ui.adsabs.harvard.edu/abs/2010ApJ...718.1266M} {718, 1266}

\bibitem[\protect\citeauthoryear{{Murray} \& {Lin}}{{Murray} \&
  {Lin}}{1992}]{Murray.Lin.1992}
{Murray} S.~D.,  {Lin} D. N.~C.,  1992, \mn@doi [\apj] {10.1086/171993}, \href
  {https://ui.adsabs.harvard.edu/abs/1992ApJ...400..265M} {400, 265}

\bibitem[\protect\citeauthoryear{{Nadler} et~al.,}{{Nadler}
  et~al.}{2020}]{Nadler.etal.2020}
{Nadler} E.~O.,  et~al., 2020, \mn@doi [\apj] {10.3847/1538-4357/ab846a}, \href
  {https://ui.adsabs.harvard.edu/abs/2020ApJ...893...48N} {893, 48}

\bibitem[\protect\citeauthoryear{{Ness}, {Asplund}  \& {Casey}}{{Ness}
  et~al.}{2014}]{ngc6522}
{Ness} M.,  {Asplund} M.,   {Casey} A.~R.,  2014, \mn@doi [\mnras]
  {10.1093/mnras/stu2144}, \href
  {https://ui.adsabs.harvard.edu/abs/2014MNRAS.445.2994N} {445, 2994}

\bibitem[\protect\citeauthoryear{{O'Malley} \& {Chaboyer}}{{O'Malley} \&
  {Chaboyer}}{2018}]{ngc6584}
{O'Malley} E.~M.,  {Chaboyer} B.,  2018, \mn@doi [\apj]
  {10.3847/1538-4357/aab554}, \href
  {https://ui.adsabs.harvard.edu/abs/2018ApJ...856..130O} {856, 130}

\bibitem[\protect\citeauthoryear{Oliphant}{Oliphant}{2015}]{NumPy}
Oliphant T.~E.,  2015, Guide to NumPy, 2nd edn.
CreateSpace Independent Publishing Platform, USA

\bibitem[\protect\citeauthoryear{{Origlia}, {Valenti}  \& {Rich}}{{Origlia}
  et~al.}{2008}]{ngc6440}
{Origlia} L.,  {Valenti} E.,   {Rich} R.~M.,  2008, \mn@doi [\mnras]
  {10.1111/j.1365-2966.2008.13492.x}, \href
  {https://ui.adsabs.harvard.edu/abs/2008MNRAS.388.1419O} {388, 1419}

\bibitem[\protect\citeauthoryear{Pedregosa et~al.,}{Pedregosa
  et~al.}{2011}]{sklearn}
Pedregosa F.,  et~al., 2011, J. Mach. Learn. Res., 12, 2825–2830

\bibitem[\protect\citeauthoryear{{Peebles}}{{Peebles}}{1965}]{Peebles.1965}
{Peebles} P.~J.~E.,  1965, \mn@doi [\apj] {10.1086/148417}, \href
  {https://ui.adsabs.harvard.edu/abs/1965ApJ...142.1317P} {142, 1317}

\bibitem[\protect\citeauthoryear{{Peebles}}{{Peebles}}{1984}]{Peebles.1984}
{Peebles} P.~J.~E.,  1984, in {Audouze} J.,  {Tran Thanh Van} J.,  eds,  NATO
  Advanced Study Institute (ASI) Series C Vol. 117, Formation and Evolution of
  Galaxies and Large Structures in the Universe. p.~185

\bibitem[\protect\citeauthoryear{{Peebles} \& {Dicke}}{{Peebles} \&
  {Dicke}}{1968}]{Peebles.Dicke.1968}
{Peebles} P.~J.~E.,  {Dicke} R.~H.,  1968, \mn@doi [\apj] {10.1086/149811},
  \href {https://ui.adsabs.harvard.edu/abs/1968ApJ...154..891P} {154, 891}

\bibitem[\protect\citeauthoryear{{Peebles} \& {Yu}}{{Peebles} \&
  {Yu}}{1970}]{Peebles.Yu.1970}
{Peebles} P.~J.~E.,  {Yu} J.~T.,  1970, \mn@doi [\apj] {10.1086/150713}, \href
  {https://ui.adsabs.harvard.edu/abs/1970ApJ...162..815P} {162, 815}

\bibitem[\protect\citeauthoryear{{Pfeffer}, {Lardo}, {Bastian}, {Saracino}  \&
  {Kamann}}{{Pfeffer} et~al.}{2021}]{Pfeffer.etal.2021}
{Pfeffer} J.,  {Lardo} C.,  {Bastian} N.,  {Saracino} S.,   {Kamann} S.,  2021,
  \mn@doi [\mnras] {10.1093/mnras/staa3407}, \href
  {https://ui.adsabs.harvard.edu/abs/2021MNRAS.500.2514P} {500, 2514}

\bibitem[\protect\citeauthoryear{{Pillepich} et~al.,}{{Pillepich}
  et~al.}{2018}]{Pilepich.etal.2018}
{Pillepich} A.,  et~al., 2018, \mn@doi [\mnras] {10.1093/mnras/stx3112}, \href
  {https://ui.adsabs.harvard.edu/abs/2018MNRAS.475..648P} {475, 648}

\bibitem[\protect\citeauthoryear{{Qu} et~al.,}{{Qu}
  et~al.}{2017}]{Qu.etal.2017}
{Qu} Y.,  et~al., 2017, \mn@doi [\mnras] {10.1093/mnras/stw2437}, \href
  {https://ui.adsabs.harvard.edu/abs/2017MNRAS.464.1659Q} {464, 1659}

\bibitem[\protect\citeauthoryear{{Rain}, {Villanova}, {Mun{\~o}z}  \&
  {Valenzuela-Calderon}}{{Rain} et~al.}{2019}]{ngc6809}
{Rain} M.~J.,  {Villanova} S.,  {Mun{\~o}z} C.,   {Valenzuela-Calderon} C.,
  2019, \mn@doi [\mnras] {10.1093/mnras/sty3208}, \href
  {https://ui.adsabs.harvard.edu/abs/2019MNRAS.483.1674R} {483, 1674}

\bibitem[\protect\citeauthoryear{{Recio-Blanco}}{{Recio-Blanco}}{2018}]{Recio-Blanco2018}
{Recio-Blanco} A.,  2018, \mn@doi [\aap] {10.1051/0004-6361/201833179}, \href
  {https://ui.adsabs.harvard.edu/abs/2018A&A...620A.194R} {620, A194}

\bibitem[\protect\citeauthoryear{{Reina-Campos}, {Trujillo-Gomez}, {Deason},
  {Kruijssen}, {Pfeffer}, {Crain}, {Bastian}  \& {Hughes}}{{Reina-Campos}
  et~al.}{2022a}]{Reina.Campos.etal2022b}
{Reina-Campos} M.,  {Trujillo-Gomez} S.,  {Deason} A.~J.,  {Kruijssen}
  J.~M.~D.,  {Pfeffer} J.~L.,  {Crain} R.~A.,  {Bastian} N.,   {Hughes} M.~E.,
  2022a, \mn@doi [\mnras] {10.1093/mnras/stac1126}, \href
  {https://ui.adsabs.harvard.edu/abs/2022MNRAS.513.3925R} {513, 3925}

\bibitem[\protect\citeauthoryear{{Reina-Campos}, {Keller}, {Kruijssen},
  {Gensior}, {Trujillo-Gomez}, {Jeffreson}, {Pfeffer}  \&
  {Sills}}{{Reina-Campos} et~al.}{2022b}]{Reina.Campos.etal.2022}
{Reina-Campos} M.,  {Keller} B.~W.,  {Kruijssen} J.~M.~D.,  {Gensior} J.,
  {Trujillo-Gomez} S.,  {Jeffreson} S. M.~R.,  {Pfeffer} J.~L.,   {Sills} A.,
  2022b, \mn@doi [\mnras] {10.1093/mnras/stac1934}, \href
  {https://ui.adsabs.harvard.edu/abs/2022MNRAS.517.3144R} {517, 3144}

\bibitem[\protect\citeauthoryear{{Rix} et~al.,}{{Rix} et~al.}{2022}]{Rix2022}
{Rix} H.-W.,  et~al., 2022, \mn@doi [\apj] {10.3847/1538-4357/ac9e01}, \href
  {https://ui.adsabs.harvard.edu/abs/2022ApJ...941...45R} {941, 45}

\bibitem[\protect\citeauthoryear{{Schweizer}}{{Schweizer}}{1987}]{Schweizer.1987}
{Schweizer} F.,  1987, in {Faber} S.~M.,  ed., Nearly Normal Galaxies. From the
  Planck Time to the Present. p.~18

\bibitem[\protect\citeauthoryear{{Searle}}{{Searle}}{1977}]{Searle.1977}
{Searle} L.,  1977, in {Tinsley} B.~M.,  {Larson} Richard B.~Gehret D.~C.,
  eds, Evolution of Galaxies and Stellar Populations. p.~219

\bibitem[\protect\citeauthoryear{{Searle} \& {Zinn}}{{Searle} \&
  {Zinn}}{1978}]{Searle.Zinn.1978}
{Searle} L.,  {Zinn} R.,  1978, \mn@doi [\apj] {10.1086/156499}, \href
  {https://ui.adsabs.harvard.edu/abs/1978ApJ...225..357S} {225, 357}

\bibitem[\protect\citeauthoryear{{Semenov}, {Conroy}, {Chandra}, {Hernquist}
  \& {Nelson}}{{Semenov} et~al.}{2023}]{Semenov.etal.2023a}
{Semenov} V.~A.,  {Conroy} C.,  {Chandra} V.,  {Hernquist} L.,   {Nelson} D.,
  2023, \mn@doi [arXiv e-prints] {10.48550/arXiv.2306.09398}, \href
  {https://ui.adsabs.harvard.edu/abs/2023arXiv230609398S} {p. arXiv:2306.09398}

\bibitem[\protect\citeauthoryear{{Simons} et~al.,}{{Simons}
  et~al.}{2017}]{Simons.etal.2017}
{Simons} R.~C.,  et~al., 2017, \mn@doi [\apj] {10.3847/1538-4357/aa740c}, \href
  {https://ui.adsabs.harvard.edu/abs/2017ApJ...843...46S} {843, 46}

\bibitem[\protect\citeauthoryear{{Sneden}, {Kraft}, {Guhathakurta}, {Peterson}
  \& {Fulbright}}{{Sneden} et~al.}{2004}]{ngc5272}
{Sneden} C.,  {Kraft} R.~P.,  {Guhathakurta} P.,  {Peterson} R.~C.,
  {Fulbright} J.~P.,  2004, \mn@doi [\aj] {10.1086/381907}, \href
  {https://ui.adsabs.harvard.edu/abs/2004AJ....127.2162S} {127, 2162}

\bibitem[\protect\citeauthoryear{{Souza} et~al.,}{{Souza}
  et~al.}{2023}]{ngc6355}
{Souza} S.~O.,  et~al., 2023, \mn@doi [\aap] {10.1051/0004-6361/202245286},
  \href {https://ui.adsabs.harvard.edu/abs/2023A&A...671A..45S} {671, A45}

\bibitem[\protect\citeauthoryear{{Spitler} \& {Forbes}}{{Spitler} \&
  {Forbes}}{2009}]{Spitler.Forbes.2009}
{Spitler} L.~R.,  {Forbes} D.~A.,  2009, \mn@doi [\mnras]
  {10.1111/j.1745-3933.2008.00567.x}, \href
  {https://ui.adsabs.harvard.edu/abs/2009MNRAS.392L...1S} {392, L1}

\bibitem[\protect\citeauthoryear{{Sun}, {Wang}, {Liu}, {Long}, {Chen}  \&
  {Gao}}{{Sun} et~al.}{2023}]{Sun.etal.2023}
{Sun} G.,  {Wang} Y.,  {Liu} C.,  {Long} R.~J.,  {Chen} X.,   {Gao} Q.,  2023,
  \mn@doi [Research in Astronomy and Astrophysics] {10.1088/1674-4527/ac9e91},
  \href {https://ui.adsabs.harvard.edu/abs/2023RAA....23a5013S} {23, 015013}

\bibitem[\protect\citeauthoryear{{Trujillo-Gomez}, {Kruijssen}, {Reina-Campos},
  {Pfeffer}, {Keller}, {Crain}, {Bastian}  \& {Hughes}}{{Trujillo-Gomez}
  et~al.}{2021}]{Trujillo_Gomez.etal.2021}
{Trujillo-Gomez} S.,  {Kruijssen} J.~M.~D.,  {Reina-Campos} M.,  {Pfeffer}
  J.~L.,  {Keller} B.~W.,  {Crain} R.~A.,  {Bastian} N.,   {Hughes} M.~E.,
  2021, \mn@doi [\mnras] {10.1093/mnras/stab341}, \href
  {https://ui.adsabs.harvard.edu/abs/2021MNRAS.503...31T} {503, 31}

\bibitem[\protect\citeauthoryear{{Trujillo-Gomez}, {Kruijssen}, {Pfeffer},
  {Reina-Campos}, {Crain}, {Bastian}  \& {Cabrera-Ziri}}{{Trujillo-Gomez}
  et~al.}{2023}]{Trujillo_Gomez.etal.2023}
{Trujillo-Gomez} S.,  {Kruijssen} J.~M.~D.,  {Pfeffer} J.,  {Reina-Campos} M.,
  {Crain} R.~A.,  {Bastian} N.,   {Cabrera-Ziri} I.,  2023, \mn@doi [arXiv
  e-prints] {10.48550/arXiv.2301.05716}, \href
  {https://ui.adsabs.harvard.edu/abs/2023arXiv230105716T} {p. arXiv:2301.05716}

\bibitem[\protect\citeauthoryear{{Valenti}, {Origlia}  \& {Rich}}{{Valenti}
  et~al.}{2011}]{ngc6624}
{Valenti} E.,  {Origlia} L.,   {Rich} R.~M.,  2011, \mn@doi [\mnras]
  {10.1111/j.1365-2966.2011.18580.x}, \href
  {https://ui.adsabs.harvard.edu/abs/2011MNRAS.414.2690V} {414, 2690}

\bibitem[\protect\citeauthoryear{{VandenBerg}, {Brogaard}, {Leaman}  \&
  {Casagrande}}{{VandenBerg} et~al.}{2013}]{VdB2013}
{VandenBerg} D.~A.,  {Brogaard} K.,  {Leaman} R.,   {Casagrande} L.,  2013,
  \mn@doi [\apj] {10.1088/0004-637X/775/2/134}, \href
  {https://ui.adsabs.harvard.edu/abs/2013ApJ...775..134V} {775, 134}

\bibitem[\protect\citeauthoryear{{Vasiliev} \& {Baumgardt}}{{Vasiliev} \&
  {Baumgardt}}{2021}]{Vasiliev2021}
{Vasiliev} E.,  {Baumgardt} H.,  2021, \mn@doi [\mnras]
  {10.1093/mnras/stab1475}, \href
  {https://ui.adsabs.harvard.edu/abs/2021MNRAS.505.5978V} {505, 5978}

\bibitem[\protect\citeauthoryear{{Wegg}, {Gerhard}  \& {Portail}}{{Wegg}
  et~al.}{2015}]{Wegg2015}
{Wegg} C.,  {Gerhard} O.,   {Portail} M.,  2015, \mn@doi [\mnras]
  {10.1093/mnras/stv745}, \href
  {https://ui.adsabs.harvard.edu/abs/2015MNRAS.450.4050W} {450, 4050}

\bibitem[\protect\citeauthoryear{{Wetzel} et~al.,}{{Wetzel}
  et~al.}{2023}]{Wetzel.etal.2023}
{Wetzel} A.,  et~al., 2023, \mn@doi [\apjs] {10.3847/1538-4365/acb99a}, \href
  {https://ui.adsabs.harvard.edu/abs/2023ApJS..265...44W} {265, 44}

\bibitem[\protect\citeauthoryear{{Whitmore} \& {Schweizer}}{{Whitmore} \&
  {Schweizer}}{1995}]{Whitmore.Schweizer.1995}
{Whitmore} B.~C.,  {Schweizer} F.,  1995, \mn@doi [\aj] {10.1086/117334}, \href
  {https://ui.adsabs.harvard.edu/abs/1995AJ....109..960W} {109, 960}

\bibitem[\protect\citeauthoryear{{Whitmore}, {Schweizer}, {Leitherer}, {Borne}
  \& {Robert}}{{Whitmore} et~al.}{1993}]{Whitmore.etal.1993}
{Whitmore} B.~C.,  {Schweizer} F.,  {Leitherer} C.,  {Borne} K.,   {Robert} C.,
   1993, \mn@doi [\aj] {10.1086/116732}, \href
  {https://ui.adsabs.harvard.edu/abs/1993AJ....106.1354W} {106, 1354}

\bibitem[\protect\citeauthoryear{{Whitmore}, {Zhang}, {Leitherer}, {Fall},
  {Schweizer}  \& {Miller}}{{Whitmore} et~al.}{1999}]{Whitmore.etal.1999}
{Whitmore} B.~C.,  {Zhang} Q.,  {Leitherer} C.,  {Fall} S.~M.,  {Schweizer} F.,
    {Miller} B.~W.,  1999, \mn@doi [\aj] {10.1086/301041}, \href
  {https://ui.adsabs.harvard.edu/abs/1999AJ....118.1551W} {118, 1551}

\bibitem[\protect\citeauthoryear{{Yong}, {Grundahl}, {Nissen}, {Jensen}  \&
  {Lambert}}{{Yong} et~al.}{2005}]{ngc6752}
{Yong} D.,  {Grundahl} F.,  {Nissen} P.~E.,  {Jensen} H.~R.,   {Lambert} D.~L.,
   2005, \mn@doi [\aap] {10.1051/0004-6361:20052916}, \href
  {https://ui.adsabs.harvard.edu/abs/2005A&A...438..875Y} {438, 875}

\bibitem[\protect\citeauthoryear{{Yong} et~al.,}{{Yong} et~al.}{2014}]{ngc7089}
{Yong} D.,  et~al., 2014, \mn@doi [\mnras] {10.1093/mnras/stu806}, \href
  {https://ui.adsabs.harvard.edu/abs/2014MNRAS.441.3396Y} {441, 3396}

\bibitem[\protect\citeauthoryear{{Zepf}, {Ashman}, {English}, {Freeman}  \&
  {Sharples}}{{Zepf} et~al.}{1999}]{Zepf.etal.1999}
{Zepf} S.~E.,  {Ashman} K.~M.,  {English} J.,  {Freeman} K.~C.,   {Sharples}
  R.~M.,  1999, \mn@doi [\aj] {10.1086/300961}, \href
  {https://ui.adsabs.harvard.edu/abs/1999AJ....118..752Z} {118, 752}

\bibitem[\protect\citeauthoryear{{Zinn}}{{Zinn}}{1985}]{Zinn.1985}
{Zinn} R.,  1985, \mn@doi [\apj] {10.1086/163249}, \href
  {https://ui.adsabs.harvard.edu/abs/1985ApJ...293..424Z} {293, 424}

\bibitem[\protect\citeauthoryear{{Zinn}}{{Zinn}}{1996}]{Zinn.1996}
{Zinn} R.,  1996, in {Morrison} H.~L.,  {Sarajedini} A.,  eds,  Astronomical
  Society of the Pacific Conference Series Vol. 92, Formation of the Galactic
  Halo...Inside and Out. p.~211

\makeatother
\end{thebibliography}

\appendix

\section{Complementing APOGEE globular cluster chemistry with literature values}
\label{app:chem_ref}

We have compiled a sample of Galactic GCs with measurements of [Mg/Fe] and [Al/Fe] available both in APOGEE DR17 and in prior spectroscopic studies. This includes NGC 362 \citep[][]{ngc362}, NGC 1851 \citep[][]{ngc1851}, NGC 2808 \citep[][]{ngc2808}, NGC 3201 \citep[][]{ngc3201}, NGC 4590 \citep[][]{ngc4590}, NGC 5272 \citep[][]{ngc5272}, NGC 6121 \citep[][]{ngc6121}, NGC 6273 \citep[][]{ngc6273}, HP 1 \citep[][]{hp1}, NGC 6388 \citep[][]{ngc6388}, NGC 6553 \citep[][]{ngc6553}, NGC 6558 \citep[][]{ngc6558},  NGC 6569 \citep[][]{ngc6569}, NGC 6715 \citep[][]{ngc6715}, NGC 6723 \citep[][]{ngc6723}, NGC 6752 \citep[][]{ngc6752}, NGC 6809 \citep[][]{ngc6809}, NGC 7089 \citep[][]{ngc7089}. In the literature, where abundance measurements are available for individual stars we calculate median [Mg/Fe] and [Al/Fe], otherwise we use published mean values.

\begin{figure*}
  \centering
  \includegraphics[width=\textwidth]{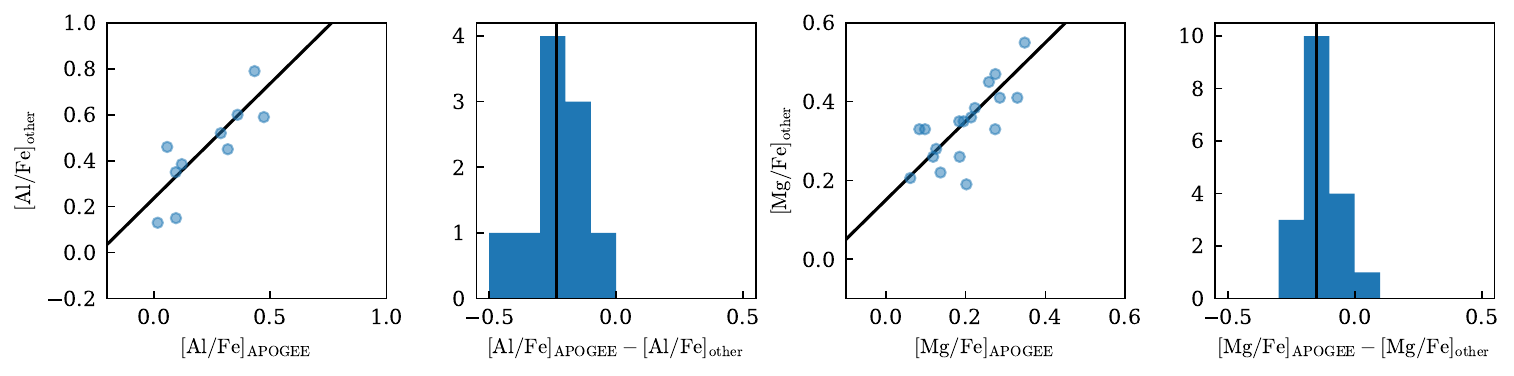}
  \caption[]{
Comparison of the chemical abundance values for several MW GCs with measurements both by APOGEE and from other previously published studies. First panel: [Al/Fe] from literature ($y$-axis) vs [Al/Fe] in APOGEE DR17 ($x$-axis). Solid line shows 1:1 relation. Second panel: Distribution of differences between [Al/Fe] values based on the APOGEE DR17 and literature. Vertical solid line shows the median offset used to place the literature values on the common scale with APOGEE. Third and Fourth panels: Same as the first two panels but for [Mg/Fe]. 
  }
  
   \label{fig:apo_lit}
\end{figure*}

Figure~\ref{fig:apo_lit} compares APOGEE DR17 ($x$ axis) and literature ($y$ axis) median/mean values of [Al/Fe] (first two panels) and [Mg/Fe] (second two panels). Compared to APOGEE DR17, literature values (based on spectroscopic studies mostly in the optical wavelength range) are higher by 0.24 dex for [Al/Fe] and by 0.15 dex for [Mg/Fe]. We subtract these constant offsets (computed as medians of the residuals for each element) from the available literature values to bring them on the same scale with APOGEE DR17.

%

\section{Distribution of accreted to in-situ fraction in simulated galaxies}
\label{app:fire_facc_elz}

%
\begin{figure*}
  \centering
  \includegraphics[width=0.9\textwidth]{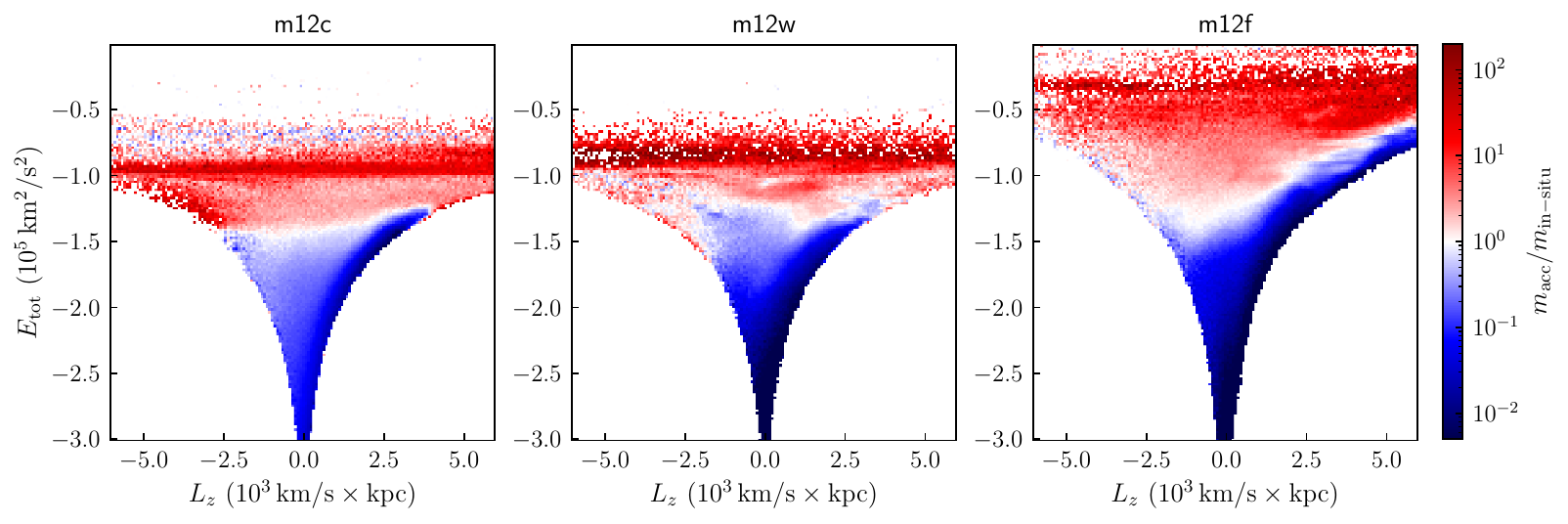}
  \includegraphics[width=0.9\textwidth]{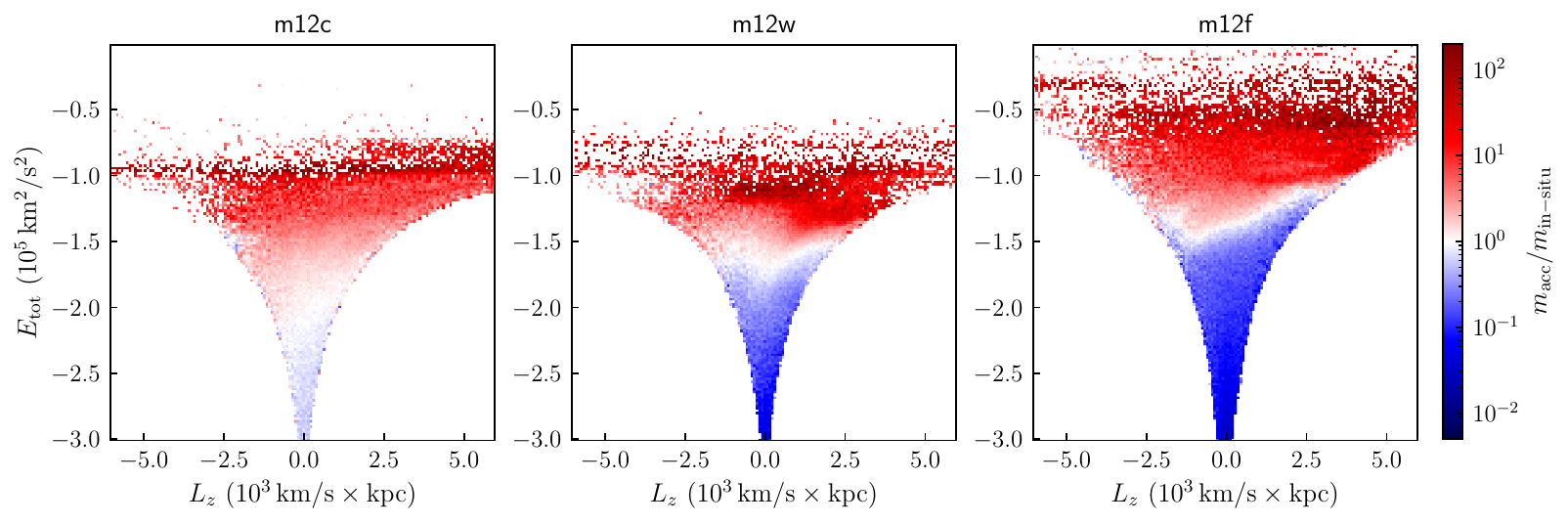}
  \caption[]{The ratio of accreted stellar mass to in-situ stellar mass, $m_{\rm acc}/m_{\rm in-situ}$,  in different regions of the total energy-angular momentum space $E-L_z$ in three MW-sized galaxies ({\tt m12c, m12w, m12f}) from the FIRE-2 suite. The {\it top row} of panels shows results for stellar particles of all metallicities, while the {\it bottom row} shows results for stellar particles with $\rm [Fe/H]<-1$ only. The color represents the logarithm of $m_{\rm acc}/m_{\rm in-situ}$, as shown on the side colormap using the divergent colormap to delineate the transition from the accretion-dominated to the in-situ dominated regions better. This boundary is delineated by the white to faint-blue color.}
   \label{fig:elz_facc}
\end{figure*}
\begin{figure}
  \centering
  \includegraphics[width=0.5\textwidth]{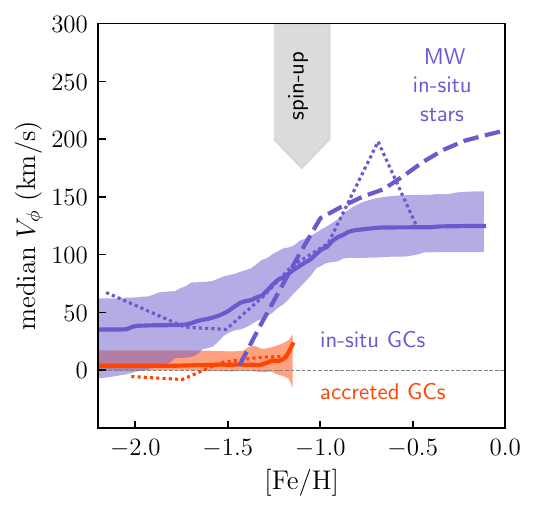}
  \caption[]{The disc spin-up traced by in-situ globular clusters similar to Figure~\ref{fig:vphifeh}, but using results of the regression fits of the functional form given by equation~\ref{eq:vphi_sigmoid} as described in the text of Appendix~\ref{app:spinup_fit}. The blue-solid lines shows median of the fits for individual GC bootstrap samples, while blue shaded region shows the region containing $68\%$ of the best fit bootstrap functions at each $\rm [Fe/H]$. As in Figure~\ref{fig:vphifeh}, blue dashed lines shows median $V_\phi$ for in-situ stars of the Milky Way, as measured in \citet{Aurora}. The red solid line and shaded region show the corresponding results for the accreted GCs. The blue and red dotted lines show medians for the bootstrap samples in the coarse bins shown in Figure~\ref{fig:vphifeh}. }
   \label{fig:spinup_fit}
\end{figure}

Figure~\ref{fig:elz_facc} shows the ratio of accreted stellar mass to in-situ stellar mass, $m_{\rm acc}/m_{\rm in-situ}$,  in different regions of the total energy-angular momentum space $E-L_z$ in three MW-sized galaxies ({\tt m12c, m12w, m12f}) from the FIRE-2 suite. The galaxies {\tt m12c} and {\tt m12w} are selected because they are close to the Milky Way in the halo and stellar mass and have the distribution of stars in the $E-L_z$ similar to the Milky Way. They also have different fractions of accreted stars.

The top row of panels shows results for stellar particles of all metallicities, while the bottom row shows results for stellar particles with $\rm [Fe/H]<-1$ only. 
The color represents the logarithm of $m_{\rm acc}/m_{\rm in-situ}$, as shown on the side colormap using the divergent colormap to delineate the transition from the accretion-dominated to the in-situ dominated regions better. This boundary is delineated by the white to faint blue color. 

Although the boundary in the top row varies from object to object in detail, reflecting different evolution pathways and merger histories, qualitatively the boundary is similar to that adopted in our classification based on the $\rm [Al/Fe]$ ratio of the MW stars. Specifically, the boundary is quite flat and is at $E\approx -1.3 \times 10^5\,\rm km^2/s^2$ at $L_z<2000\ \rm kpc\, km/s$ and increases in energy with increasing $L_z$ at $L_z>2000\ \rm kpc\, km/s$.  

Comparing bottom and top panels for simulations {\tt m12c} and {\it m12w} shows that the boundary between the accretion and in-situ dominated regions in the $E-L_z$ plane can depend on metallicity. However, we note that for the MW analysis carred out in \citet{Belokurov.Kravtsov.2023} and in this work the boundary is actually calibrated most reliably at the metallicities of $-1.5\lesssim\rm [Fe/H]\lesssim -1$.

\section{disc spin-up with globular clusters using fitting instead of binning}
\label{app:spinup_fit}

%
As an alternative to binning and estimating median and its uncertainty using coarse bins, as was done in Section~\ref{sec:gcspinup}, we model the trend of $V_\phi$ with $x=\rm [Fe/H]$ using the sigmoid function that has the shape of a ``soft step'':
\begin{equation}
s(x) = \frac{V_{\rm high}}{\exp\left(-[10x + x_{\rm sp}]\right)+1} + V_{\rm low},
    \label{eq:vphi_sigmoid}
\end{equation}
where $V_{\rm low}$ and $V_{\rm high}$ are smallest velocity at metallicities below the spin-up and $V_{\rm high}$ is the velocity increase from $V_{\rm low}$ to the maximal velocity at metallicities larger than the spin-up $\rm [Fe/H]$. The bias parameter $x_{\rm sp}$ determines the metallicity at which spin-up occurs. The factor $10$ in equation~\ref{eq:vphi_sigmoid} controls the width of the step and was fixed in the fits to minimize degeneracies between parameters. 

Specifically, we carry out the minimal absolute distance regression using metallicities and $V_\phi$ values for individual GCs and find the best-fit parameters $V_{\rm low}$, $V_{\rm high}$, $x_{\rm sp}$ minimizing the cost function:
\begin{equation}
   C = \sum\limits_{i=1}^{N_{\rm GC}}\vert s(x) - x_i\vert. 
\end{equation}
This type of regression approximates the median trend of the data points. 

We carry out such regression for 1000 bootstrap resamples of the original GC samples and plot the median and 68\% range of the best-fit functional fits to the bootstrap samples of the in-situ and accreted samples as solid lines in Figure~\ref{fig:spinup_fit}. Note that we only use accreted clusters with $\rm [Fe/H]<-1$ in the fit as there are only 3 accreted clusters in our classification at higher metallicities, which makes the fit unconstrained at these higher metallicities. The figure compares results obtained by this method with the medians obtained using bootstrap samples in coarse bins shown in Figure~\ref{fig:vphifeh} and shows that both methods produce similar results.
We conclude therefore that detection of spin-up at $\rm [Fe/H]\approx -1.3\div-1$ in in-situ GCs is robust.

\section{List of GC classifications}
\label{app:class_tab}

Table~\ref{tab:class} presents the list of globular clusters used in this study and their classification using our method (1 is in-situ, 0 is accreted). The last column provides comments for individual clusters that may be misclassified by this method, where NSC stands for the Nuclear Star Cluster based on the analysis of \citet{Pfeffer.etal.2021} and where we used analysis of element abundance ratios presented in Section~\ref{sec:prevcomp} to indicate the in-situ (accreted) clusters with the ratios similar to those of accreted (in-situ) systems.

\begin{table*}
\centering
\caption{The names of the globular clusters used in this study and their classification using our method (1 is in-situ, 0 is accreted).}
\begin{tabular}[t]{lcp{3cm}} 
    \hline
    Cluster name & In-situ/accreted (1/0) & Comments \\
    \hline
2MASS-GC01      & 1 & \\
2MASS-GC02      & 1 & \\
AM 1            & 0 & \\
AM 4            & 0 & \\
Arp 2           & 0 & \\
BH 140          & 1 & \\
BH 261          & 1 & \\
Crater          & 0 & \\
Djorg 1         & 1 & \\
Djorg 2         & 1 & \\
E 3             & 1 & \\
ESO 280-SC06    & 0 & \\
ESO 452-SC11    & 1 & \\
Eridanus        & 0 & \\
FSR 1716        & 1 & \\
FSR 1735        & 1 & \\
FSR 1758        & 0 & \\
Gran 1          & 1 & \\
Gran 2          & 1 & \\
Gran 3          & 1 & \\
Gran 5          & 1 & \\
HP 1            & 1 & \\
IC 1257         & 0 & \\
IC 1276         & 1 & \\
IC 4499         & 0 & \\
Laevens 3       & 0 & \\
Liller 1        & 1 & \\
Lynga 7         & 1 & \\
NGC 104         & 1 & \\
NGC 1261        & 0 & \\
NGC 1851        & 0 & \\
NGC 1904        & 0 & \\
NGC 2298        & 0 & \\
NGC 2419        & 0 & \\
NGC 2808        & 0 & \\
NGC 288         & 0 & [Al/Fe] and [Mg/Fe] consistent with in-situ, see Figure~\ref{fig:ecc_mg_al}\\
NGC 3201        & 0 & \\
NGC 362         & 0 & \\
NGC 4147        & 0 & \\
NGC 4372        & 1 & \\
NGC 4590        & 0 & \\
NGC 4833        & 1 & \\
NGC 5024        & 0 & \\
NGC 5053        & 0 & \\
NGC 5139        & 1 & likely NSC/accreted\\
NGC 5272        & 0 & \\
NGC 5286        & 0 & [Al/Fe] and [Mg/Fe] consistent with in-situ, see Figure~\ref{fig:ecc_mg_al}\\
NGC 5466        & 0 & \\
NGC 5634        & 0 & \\
NGC 5694        & 0 & \\
NGC 5824        & 0 & \\
NGC 5897        & 1 & \\
NGC 5904        & 0 & \\
NGC 5927        & 1 & [Al/Fe] and [Mg/Fe] consistent with being accreted, see Figure~\ref{fig:ecc_mg_al}\\
NGC 5946        & 1 & \\
NGC 5986        & 1 & \\
NGC 6093        & 1 & \\
\hline
\end{tabular}
\begin{tabular}[t]{lcp{3cm}}
\hline
Cluster name & In-situ/accreted (1/0) & Comments \\
\hline
NGC 6101        & 0 & \\
NGC 6121        & 1 & \\
NGC 6139        & 1 & \\
NGC 6144        & 1 & \\
NGC 6171        & 1 & \\
NGC 6205        & 1 & \\
NGC 6218        & 1 & \\
NGC 6229        & 0 & \\
NGC 6235        & 1 & \\
NGC 6254        & 1 & \\
NGC 6256        & 1 & \\
NGC 6266        & 1 & \\
NGC 6273        & 1 & likely NSC/accreted\\
NGC 6284        & 1 & \\
NGC 6287        & 1 & \\
NGC 6293        & 1 & \\
NGC 6304        & 1 & \\
NGC 6316        & 1 & \\
NGC 6325        & 1 & \\
NGC 6333        & 1 & \\
NGC 6341        & 0 & \\
NGC 6342        & 1 & \\
NGC 6352        & 1 & \\
NGC 6355        & 1 & [Al/Fe] and [Mg/Fe] consistent with being accreted, see Figure~\ref{fig:ecc_mg_al}\\
NGC 6356        & 1 & \\
NGC 6362        & 1 & \\
NGC 6366        & 1 & \\
NGC 6380        & 1 & \\
NGC 6388        & 1 & [Al/Fe] and [Mg/Fe] consistent with being accreted, see Figure~\ref{fig:ecc_mg_al}\\
NGC 6397        & 1 & \\
NGC 6401        & 1 & \\
NGC 6402        & 1 & \\
NGC 6426        & 0 & \\
NGC 6440        & 1 & \\
NGC 6441        & 1 & \\
NGC 6453        & 1 & \\
NGC 6496        & 1 & \\
NGC 6517        & 1 & \\
NGC 6522        & 1 & \\
NGC 6528        & 1 & [Al/Fe] and [Mg/Fe] consistent with being accreted, see Figure~\ref{fig:ecc_mg_al}\\
NGC 6535        & 1 & \\
NGC 6539        & 1 & \\
NGC 6540        & 1 & \\
NGC 6541        & 1 & \\
NGC 6544        & 1 & \\
NGC 6553        & 1 & \\
NGC 6558        & 1 & \\
NGC 6569        & 1 & \\
NGC 6584        & 0 & [Al/Fe] and [Mg/Fe] consistent with in-situ, see Figure~\ref{fig:ecc_mg_al}\\
NGC 6624        & 1 & \\
NGC 6626        & 1 & \\
NGC 6637        & 1 & \\
NGC 6638        & 1 & \\
NGC 6642        & 1 & \\
NGC 6652        & 1 & \\
\hline
\end{tabular}
\label{tab:class}
\end{table*}
\setcounter{table}{0}
\begin{table}
\centering
\caption{(continued). The names of the globular clusters used in this study and their classification using our method (1 is in-situ, 0 is accreted). The last column provides comments for individual clusters that may be misclassified by this method. NSC = Nuclear Star Cluster based on the analysis of \citet{Pfeffer.etal.2021}.}
\begin{tabular}[!ht]{lcp{3cm}} 
    \hline
    Cluster name & In-situ/accreted (1/0) & Comments \\
    \hline
NGC 6656        & 1 & \\
NGC 6681        & 1 & \\
NGC 6712        & 1 & \\
NGC 6715        & 0 & \\
NGC 6717        & 1 & \\
NGC 6723        & 1 & \\
NGC 6749        & 1 & \\
NGC 6752        & 1 & \\
NGC 6760        & 1 & \\
NGC 6779        & 0 & \\
NGC 6809        & 1 & \\
NGC 6838        & 1 & \\
NGC 6864        & 0 & \\
NGC 6934        & 0 & \\
NGC 6981        & 0 & \\
NGC 7006        & 0 & \\
NGC 7078        & 1 & Shows enhanced [Eu/Fe] and may be accreted (see Monty et al in prep)\\
NGC 7089        & 0 & \\
NGC 7099        & 1 & \\
NGC 7492        & 0 & \\
Pal 1           & 1 & \\
Pal 10          & 1 & \\
Pal 11          & 1 & \\
Pal 12          & 0 & \\
Pal 13          & 0 & \\
Pal 14          & 0 & \\
Pal 15          & 0 & \\
Pal 2           & 0 & \\
Pal 3           & 0 & \\
Pal 4           & 0 & \\
Pal 5           & 0 & \\
Pal 6           & 1 & \\
Pal 8           & 1 & \\
Patchick 126    & 1 & \\
Pyxis           & 0 & \\
Rup 106         & 0 & \\
Sagittarius II  & 0 & \\
Terzan 1        & 1 & \\
Terzan 10       & 1 & \\
Terzan 12       & 1 & \\
Terzan 2        & 1 & \\
Terzan 3        & 1 & \\
Terzan 4        & 1 & \\
Terzan 5        & 1 & \\
Terzan 6        & 1 & \\
Terzan 7        & 0 & \\
Terzan 8        & 0 & \\
Terzan 9        & 1 & \\
Ton 2           & 1 & \\
UKS 1           & 1 & \\
VVV-CL001       & 1 & \\
VVV-CL160       & 0 & \\
Whiting 1       & 0 & \\
\hline
\end{tabular}
\end{table}

\label{lastpage}

\end{document}